\RequirePackage{rotating} 
\documentclass{svjour3}
\usepackage[dvipsnames,table]{xcolor}
\usepackage{graphicx}
\usepackage{subfig}

\usepackage{scrextend}
\usepackage{url}
\usepackage{doi} 
\usepackage{hyperref}

\usepackage{amsmath,amssymb,latexsym}
\usepackage{mathabx}

\usepackage{tikz}
\usepackage{pgfplots}
\pgfplotsset{compat=1.17}
\usetikzlibrary{shapes,arrows}
\usepgfplotslibrary{statistics}
\usepackage{bchart}
\usepackage{tabularx}
\usepackage{booktabs}
\usepackage{caption}
\usepackage[para,online,flushleft]{threeparttable}
\usepackage{array}
\usepackage{makecell}

\usepackage{siunitx}
\usepackage{numprint}
\npstyleenglish

\usepackage{rotating} 

\usepackage{enumitem}
\usepackage[shortcuts]{extdash}

\usepackage{xcolor}
\usepackage{ifthen}
\newboolean{showcomments}
\setboolean{showcomments}{false} 
\ifthenelse{\boolean{showcomments}}
    {
        \newcommand{\rasmus}[1]{\textcolor{gray}{{\it [Rasmus says: #1]}}}
        \newcommand{\elizabeth}[1]{\textcolor{red}{{\it [Elizabeth says: #1]}}}
        \newcommand{\per}[1]{\textcolor{blue}{{\it [Per says: #1]}}}
    }
    {
        \newcommand{\rasmus}[1]{}
        \newcommand{\elizabeth}[1]{}
        \newcommand{\per}[1]{}
    }

\begin{document}

\title{A Theory of Factors Affecting Continuous Experimentation (FACE)}

\author{Rasmus Ros \and Elizabeth Bjarnason \and Per Runeson}
\authorrunning{Ros et al.}

\institute{
    Rasmus Ros 
    \at Lund University, Sweden \\
    \email{rasmus.ros@cs.lth.se}
\and
    Elizabeth Bjarnason
    \at Lund University, Sweden \\
    \email{elizabeth.bjarnason@cs.lth.se}
\and
    Per Runeson
    \at Lund University, Sweden \\
    \email{per.runeson@cs.lth.se}
}

\date{Received: date / Accepted: date}

\maketitle

\begin{abstract}
\emph{Context:}
Continuous experimentation (CE) is used by many companies with internet-facing products to improve their software based on user data. Some companies deliberately adopt an experiment-driven approach to software development while some companies use CE in a more ad-hoc fashion.
\emph{Objective:}
The goal of the study is to identify factors that explain the variations in the utility and efficacy of CE between different companies.
\emph{Method:}
We conducted a multi-case study of 12 companies involved with CE and performed 27 interviewees with practitioners at these companies. Based on that empirical data, we then built a theory of factors at play in CE.
\emph{Results:}
We introduce a theory of Factors Affecting Continuous Experimentation (FACE). The theory includes three factors, namely 1) processes and infrastructure for CE, 2) the user problem complexity of the product offering, and 3) incentive structures for CE. It explains how these factors affect the effectiveness of CE and its ability to achieve problem-solution and product-market fit.
\emph{Conclusions:}
Our theory can be used by practitioners to assess an organisation's potential for adopting CE, as well as, identifying  factors which pose challenges in gaining value from CE practices. Our results also provide a starting point for further research on how contextual factors affect CE and how these may be mitigated.
\end{abstract}

\section{Introduction}
\label{sec:introduction}

Continuous experimentation (CE) is an experiment-driven software engineering approach where assumptions about product features and requirements are continuously tested with users. The aim is to reduce the risk of wasting development resources on requirements of little or no value to the users. In CE, software changes are compared to an old version in an experiment (such as an A/B test) with real users and only changes with a positive effect on usage are permanently released. Initially, this approach was primarily applied by large internet-facing software engineering companies such as Microsoft~\cite{kohavi2009online,kohavi2009controlled}, Google~\cite{tang2010overlapping}, and Facebook~\cite{feitelson2013development} that apply an agile and continuous approach to software engineering. Lately, also more traditional software engineering organizations such as companies in the business-to-business domain~\cite{rissanen2015continuous} are recognising the value of moving towards an experiment-driven development model that provides the ability to evaluate product assumptions through continuous user feedback~\cite{fagerholm2017right}.

Performing CE is not an easy task and one that requires companies to adapt their processes, structures, technical infrastructures, and culture~\cite{fagerholm2014building,yaman2017introducing}. Several researchers report on the benefits of adopting CE and a high-level model on how CE is structured (the RIGHT model) has been proposed~\cite{fagerholm2017right}. Despite this, not all companies are able to adopt CE, even though they might have sufficient user data available. We pose that this is due to the complex nature of software engineering and that many different factors and processes are at play. Some of these factors have been identified in previous research, such as having a service-oriented offering~\cite{schermann2018we} or one with cyber-physical systems~\cite{bosch2012eternal,giaimo2016continuous}, and whether the target users are consumers (B2C) or other businesses (B2B)~\cite{rissanen2015continuous,yaman2016transitioning}. Also, only some companies are able to conduct experiments that affect business value~\cite{schermann2018we}. 
However, there is a lack of overarching theory that describes which factors that are at play and that explains how these affect the efficacy of CE for different contexts. For this reason, we wanted to investigate if there are any common patterns that can explain the varying utility, challenges, and benefits~\cite{auer2021controlled} experienced by companies when applying CE.

We have performed a multi-case study of 12 companies to explore CE practices and contexts~\cite{clarke2012situational,petersen2009context} for a range of companies, and constructed an empirically-based theory through iterative and systematic analysis of this data. Our empirical data consists of 27 semi-structured interviews with practitioners working in various roles relevant to CE, such as software developers, quality assurance, data scientists, and product owners. We have previously published two initial findings based on part of the interview material. First, a paper with five interviews about scenarios in which experiments are used~\cite{ros2018continuous} which give an indication that the context in which CE is performed matters. Second, a paper with 14 interviews on the role of business models with product-led growth~\cite{ros2020continuous}. In this paper, we expand on the initial findings by including seven additional case companies, and by deepening the analysis and generating a theory based on the full set of interviews.

We present a theory of Factors Affecting Continuous Experimentation (FACE) based on empirical data from 12 case companies. FACE considers socio-technical factors in organizational contexts surrounding CE and observed relationships between factors. According to FACE, there are three main factors that influence how well an organization can expect to conduct CE. These factors are: (1) \emph{CE processes and infrastructure}, in particular, the data infrastructure that companies have for conducting telemetry and analysis of results; (2) the complexity of the \emph{problem} that the software solves for its users limits the applicability of CE. For example, a software system that integrates many other systems is highly complex and difficult to make changes in, thereby hindering CE; and (3) \emph{incentives} related to the business model and product improvements that affect the ease of defining relevant metrics for CE and thereby influence an organisations ability to adopt CE. 


The rest of this paper is structured as follows.
The background and related work on CE and theory building are presented in Section~\ref{sec:background}, and our research method is outlined in Section~\ref{sec:method}. Our FACE theory is presented in Section~\ref{sec:theory} and the empirical underpinnings and explanations for the theory are provided in Section~\ref{sec:results}. The findings are discussed in Section~\ref{sec:discussion} including limitations and factors of CE. Finally, we conclude the study in Section~\ref{sec:conclusions}.
\section{Background and Related Work}
\label{sec:background}

Companies in many different sectors have adopted Continuous Experimentation (CE)~\cite{auer2018current,ros2018continuous2}, where features are evaluated through user feedback. Prototypes of a feature or product can be quickly validated with users before a costly implementation is finalized and released to all users. After implementation, the change in software can be subjected to a controlled experiment (such as an A/B test) where a comparison can be made with and without the new change. Only changes that have a positive impact on user feedback are accepted. The results of an experiment might beget further questions, especially for negative results, to figure out what went wrong. Thus, experiments are usually executed in a sequence, which is why the practice is coined \emph{continuous experimentation}. Both pre-deployment experiments (performed during prototyping) and controlled post-deployment experiments are of interest in this study. The literature topics covered in this section include software business model and how it relates to CE, an overview and related work of CE, and theory building in software engineering.

\subsection{Software business models}
\label{sec:background_business_models}

CE provides a way to measure the value that software development brings to users and business. As such, studying CE entails understanding how that value is delivered. In this study, we use business models and business strategy as a lens to structure our analysis of that value delivery. A \emph{business strategy} in the management field is a long-term vision for a company~\cite{johnson2008exploring}. A business model is a concrete plan to execute that vision. The term \emph{business model} is narrowly used in industry to refer to how a company collects revenue~\cite{vanhala2013we}. In this paper, the broader definition by Osterwalder~\cite{osterwalder2010business} is used: ``\emph{A business model describes the rationale of how an organization creates, delivers, and captures value}''.

In the business model innovation field, prototype experiments have been studied as a method to find a combination of a working business model and product~\cite{brunswicker2013business,sorescu2017data,wrigley2016designing}. This contrasts with the way that experiments are used in CE research, where the focus is more on product improvement~\cite{fagerholm2017right}, and changing the business model might be considered out of scope for daily software engineering work. In addition, successfully implementing changes in an established business model is notoriously hard and risky~\cite{chesbrough2007business,chesbrough2002role}. As such, business models are mainly used to describe the context of companies in this study, not as a subject of experiments.

There have been several attempts to describe commonalities in software engineering business models with frameworks~\cite{rajala2003framework,schief2012business}. For example, Rajala et al.~\cite{rajala2003framework} include aspects of product strategy, revenue logic, distribution model, and service and implementation model. The business model canvas~\cite{osterwalder2004business} is probably the most popular framework to describe business models succinctly. We use an adaption of this framework as described in the next subsection.

\subsubsection{Lean startup}
\label{ref:background_lean}

Lean startup is a methodology, originating in industry from Ries~\cite{ries2011lean}, that applies lean manufacturing principles~\cite{krafcik1988triumph} to entrepreneurship in general. Lean startup has also been studied in a software engineering context~\cite{bjarnason2021prototyping,bosch2013early,fagerholm2017right}. The idea is to conduct product development in short cycles to obtain feedback on whether a proposed business model is viable as early as possible. Ries calls it the build--measure--learn cycle, where each cycle consist of a business hypothesis and an experiment to verify the hypothesis. As such, CE and lean startup have a clear connection.

The kind of CE that lean startup calls for is done through prototyping a solution and testing it on a limited set of customers~\cite{bjarnason2021prototyping,gutbrod2017software,vargas2020understanding}. The goal is then to find a minimum viable product (MVP), which is the smallest set of features that solve the users' problems. It cannot give accurate numbers that can be used to compare different solutions, as is the case with A/B testing. As such, the prototyping experiments have low fidelity but have a low cost compared to A/B tests that must be executed on finalized software in a production environment.

Maurya proposes an adaptation of the business model canvas to suit lean startup needs, called the \emph{lean canvas}~\cite{maurya2012running}. The lean canvas is divided into a product and a market part. The product part contains: the problem--solution pair that the product addresses, what key metrics are measured, and what the cost structure is for acquiring customers, developing code, operations, etc. The market part contains what the unfair advantage is, such that the product cannot be easily copied, what the channels to customers are, what the target customers are, and the revenue streams. The two parts are tied together with a value proposition message.

Maurya~\cite{maurya2012running} also describes the three phases of a startup that (1) start with finding a working problem--solution pair, (2) then the product should have a market fit, and finally (3) once this is validated can growth be the focus. Experiments are used differently at these stages where the initial focus is prototyping and later on controlled experiments can tune the product for product--market fit and growth.

\subsubsection{Product-led growth and growth engineering}
\label{sec:background_growth}

A specific archetype of business models has recently been popularized in industry, under the name of \emph{product-led growth}~\cite{bartlett2020what}. A business model that has a product-led growth relies on the product itself to acquire new end-users rather than on the direct sales \& marketing activities (e.g. advertisement or cold calling) of the sales-led business models. The purpose is to have an offering that can scale to high levels of demand. Furthermore, a new role has also been introduced to business with product-led growth~\cite{kemell2019software,troisi2020growth}, called growth marketers, growth engineers, or growth hackers. These roles are hybrids between marketers and software engineers that work with experiments in a data-driven fashion to propel a company's customer acquisition growth. Two of the companies in our study have  employees with such titles.

A more precise definition of product-led growth---as the term is used in industry---has not been found. Instead the following characteristics are derived from Bartlett~\cite{bartlett2020what}:
\begin{itemize}
    \item the software development organization elicits requirements in order to meet market needs;
    \item there is no customer specific development in order to ensure software development is directed towards improving the product for all users;
    \item the channels to acquire customers are scalable to many customers and are often organic (i.e. word of mouth instead of direct sales);
    \item the primary source of revenue is through product sales or subscriptions.
\end{itemize}

The last of the above characteristic, regarding the source of revenue, is related to the licensing model that software is sold under. According to Bartlett~\cite{bartlett2020what}, product-led growth is associated with \emph{freemium}. A freemium product~\cite{niculescu2011should} is available both for free and as a paid premium version. The premium version might, for example, have more features or offer improved customer support. As such, freemium encourages growth by allowing more customers to use the product and spread the word.

\subsection{Continuous experimentation}
\label{sec:background_ce}

CE has been studied from many different perspectives~\cite{ros2018continuous}. In software engineering venues, the topics have been varied, e.g., designs of specialized tools for optimizing experiments~\cite{ros2020data,schermann2018search} and descriptions of the CE process in use at various companies~\cite{bosch2012building,fagerholm2017right}. Although software engineering is the focus of this study, there has also been considerable practitioner focused research on CE in data science and user experience research venues. In the data science field, the seminal paper by Kohavi et al.~\cite{kohavi2009controlled} provides a practical guide to starting with controlled experiments on software in a production environment. In user experience research~\cite{sauro2016quantifying,schumacher2009handbook}, CE has a less prominent role among other competing methods such as user observations, card sorting, or interviews.

\subsubsection{Process and infrastructure models}
\label{sec:background_models}

The infrastructure and process of how several different companies conduct CE have been described with reference models and experience reports~\cite{bosch2012building,feitelson2013development,kohavi2009online}. The models are conceptual generalizations based on observations of CE in industry, thus they cannot be used to explain organizations' efficacy of CE which is the purpose of FACE. The RIGHT model by Fagerholm et al.~\cite{fagerholm2017right} contains a description of the process and infrastructure needs for conducting CE based on a multi-case study. The earlier HYPEX model by Holmstr{\"o}m Olsson and Bosch~\cite{olsson2014hypex} has similar goal but is less comprehensive. These reports and models served as a starting point for the questions in the interview guide in this study, in particular the RIGHT model.

The process model in RIGHT is inspired by the build--measure--learn cycle of lean start-up~\cite{ries2011lean}. Fagerholm et al. use the concepts of a minimum viable feature (MVF) to bridge the theoretical gap between prototyping experiments and controlled experiments, and so the model considers both types of experiments. There are five main phases of the CE process in RIGHT. (1)~In the \emph{ideation} phase hypotheses are elicited and prioritized and a change to the software is proposed. (2)~\emph{Implementation} of the minimum change that tests the hypothesis follows. (3)~Then, a suitable \emph{experiment design} and a criterion for success is selected (a metric in the case of a controlled experiment). (4)~\emph{Execution} involves deploying the product into production and monitoring the experiment. Finally, (5)~an \emph{analysis} and decision is made whether the results are satisfactory; if not the process restarts.

The technical infrastructure needs in RIGHT include tools for managing experiments and analytics, instrumentation in the product, and a continuous delivery pipeline. These tools are often referred to as an \emph{experimentation platform} when considered as a whole~\cite{gupta2018anatomy,kohavi2009controlled}. The organizational infrastructure includes roles involved with CE, of which there are many, since the CE phases cover the whole software engineering process. The necessary roles are according to Fagerholm et al.~\cite{fagerholm2017right}: Business Analysts and Product Owners elicit hypotheses and maintain a CE road map; Data Scientists design, execute, and analyze experiments; Software Developers and Quality Assurance develop and verify the software; and Operations Engineers and Release Engineers deploy and deliver the software. 

\subsubsection{Factors affecting continuous experimentation}
\label{sec:background_factors}

We are aware of one other attempt at analysis of how effective CE is for various companies, albeit from a startup perspective. Melagati et al.~\cite{melegati2020xpro} studied factors affecting CE in terms of enablers and inhibitors of CE at early-stage startups. Many of the identified inhibitors point to a lack of resources to conduct experiments, which is more pressing for startups due to a general lack of resources. Also, the research only considers pre-deployment prototype experiments, which is much less technically demanding.

There are also many studies that describe or report experiences about conducting CE for various circumstances. We have identified five such clusters of papers with domain specific challenges in a systematic literature review~\cite{auer2021controlled}. Two of the clusters relate to factors that might influence the gains with CE: business-to-business (B2B) and cyber-physical systems. The challenges in the remaining three clusters are either overlapping (mobile and cyber-physical systems) or potentially solved by statistical solutions (e-commerce and social media).

The challenges with CE in the \emph{business-to-business} (B2B) cluster are many~\cite{rissanen2015continuous,ros2018continuous,yaman2016transitioning}. B2B companies are usually involved with their customers' software engineering or IT departments, which causes issues with control of software deployment or with access to end-user data. Both of these are necessary and require constant collaboration efforts to resolve. Furthermore, the incentives to improve the software product in terms of end-user experience might not be there, for example, if the company in question generates revenue per project instead of through the value delivered to users. These issues might not be faced to the same degree by companies with a business-to-consumer (B2C) business model. Whether a company is B2B or B2C is not an explicit part of FACE; these challenges are instead explained by other constructs in FACE (incentives driven by business model and problem complexity).

As for the \emph{cyber-physical systems}~\cite{bosch2012eternal,giaimo2016continuous,mattos2018challenges,olsson2019data} cluster, the focus of the research is on suggesting and describing the required infrastructure to enable CE, such as continuous delivery of software and how telemetry can be implemented. Without this, CE is not possible. The body of work on these domains is still in the early stages and no such companies are included in this study.

\subsubsection{Previous work}
\label{sec:background_previous}

We have two prior publications from a research project started in year 2018 that uses five of the 12 cases that this study is based on~\cite{ros2018continuous,ros2020continuous}. The findings from these papers are synthesized and expanded  in FACE. 

The first paper~\cite{ros2018continuous} was a single case study on a B2B company that develops behavior algorithms and used CE in four different scenarios. (1) To verify that changes in their algorithms are beneficial. (2) To help their customers use their algorithms correctly. (3) To prove that their algorithms outperform their competitors'. Finally, (4) as a black box optimization method to tune the algorithms automatically~\cite{broden2019bandit}. The degree of tool support was different between the scenarios, ranging from fully automated to no support. The scenarios give an early indication on how CE differs depending on the purpose.

The second paper~\cite{ros2020continuous} used five of the 12 cases in a comparative case study. The study was about the role that a company's business model plays in relation to CE. The differences attributed to business models were sufficiently explained by whether the business model had product-led growth or not (corresponding to one of the three factors derived from FACE). Four drivers of a product-led growth focus were identified as affecting CE. (1)~Development and sales \& marketing worked more closely together in the product-led cases, with mutual benefits. (2)~The prioritization process used data to a higher degree to inform development. (3)~What features were included were based on market needs rather than by customer requests, thereby decreasing excessive feature bloat. Finally, (4)~the availability of metrics relevant to business was higher in the product-led cases.

\subsection{Theory building}
\label{sec:background_theory}

Theories provide a means of structuring and conveying knowledge in a condensed form. Theories can be of different types, some provide explanations of phenomena or predict the outcome, e.g., of applying a certain practice. Theories that describe a particular aspect of software engineering are gaining traction (see, e.g., Rodriguez et al.~\cite{rodriguez2020theory} or Munir et al.~\cite{munir2018theory}), presumably due to their usefulness in supporting research design and on improving software engineering practice. According to Stol and Fitzgerald~\cite{stol2015theory}, it is preferable to base practitioner guidelines on theories such that they are underpinned with theoretical knowledge on why they hold.


We used the guidelines by Sj{\o}berg et al.~\cite{sjoberg2008building} to build FACE. In Sj{\o}berg et al., the theory is constructed based on a generalization of multiple cases observed in real world. There are other theory building approaches~\cite{wieringa2015six}, such as basing them on existing theory from other fields or grounded theory~\cite{glaser1967Discovery,stol2016grounded}, which is used when the researchers want no preconceived notions about how the data should be interpreted. A theory, as defined by Sj{\o}berg et al.~\cite{sjoberg2008building}, consists of \emph{constructs} that the theory makes statements about in the form of \emph{propositions}, which are relations between the constructs. The theory also contains explanations about why the propositions hold and a scope that the theory is valid under.

\section{Method}
\label{sec:method}

\begin{figure}[t]
    \centering
    \scriptsize
    \textsf{
    \input{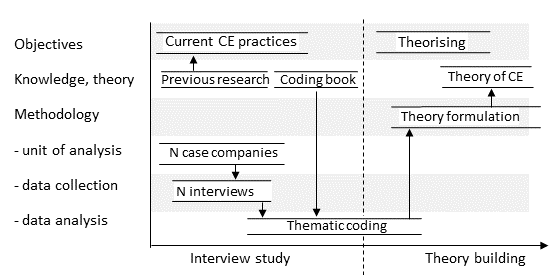}
    }
    \caption{An overview of the research method through which FACE was constructed based on a multi-case study. Two previous studies (to the left) were used as a starting point for the multi-case study.}
    \label{fig:Rmethod}
\end{figure}

This paper presents a \emph{multi-case study} of companies applying CE and a \emph{theory building} process aimed at providing a generalized description of factors affecting a company's ability to obtain gains from CE. The resulting theory was inducted through iterative analysis of the interview data. An overview of our research method is provided in Figure~\ref{fig:Rmethod}. The empirical data in our study is from case companies and we use the guidelines by Runeson et al.~\cite{runeson2009guidelines} for case study research, covering the interview process and thematic coding. The theory was then created through the codes and themes, using the process for building theories by Sj{\o}berg et al.~\cite{sjoberg2008building}. The theory is based on data from 12 case companies, denoted A--L, see Section~\ref{sec:method_cases}.

\subsection{Multi-case study with interviews}
We performed \emph{semi-structured interviews} in order to gain insights into what and how contextual factors affect an organization's ability to perform CE, and thereby provide a rich empirical basis for our theory building process. We designed an \emph{interview guide} based on previous related research~\cite{fagerholm2017right,kohavi2009controlled,olsson2014hypex}, \emph{sampled and recruited companies and practitioners} to include in our study, performed the \emph{interviews} of these using an interview guide, and \emph{analyzed} the interview data using a code book. The first author led the interview study, including the design of the interview guide and the code book, recruiting of the interviewees, performing and analysing all of the interviews. The second and third authors reviewed and provided feedback on the research design and the research artefacts, provided feedback on the interview guide, participated in two interviews each, and performed independent coding of one interview to improve the coding process.

This study is part of a three year research project with two previous publications (see Section~\ref{sec:background_previous}). The project was started with an exploratory pilot case study on a company with established CE practices. Some of the findings from that pilot study has been previously published~\cite{ros2018continuous}. An initial version of the interview guide and code book used in this study was created for this pilot study. The pilot study was followed up with a first iteration of this study, on the role of business models in CE~\cite{ros2020continuous}. The first iteration study was conducted after all interviews were conducted, but included only five of the 12 cases.

An \emph{interview guide} was used to support the interviews and to ensure that all relevant aspects were covered in each interview. We designed the guide based on our knowledge of the area and of previous research on CE, in order to cover as many relevant aspects as possible, and thereby further enhance the richness of the resulting empirical data. The descriptive models by Fagerholm et al.~\cite{fagerholm2017right} and Holmstr{\"o}m and Bosch~\cite{olsson2014hypex} were used as the theoretical basis to derive the interview questions. Additional descriptions of CE from experience reports~\cite{kohavi2009online,bosch2012building} were used to ensure that the questions covered all aspects of the CE process and infrastructure, and the context around CE in terms of the organization and business. The interview guide was designed iteratively by the authors with small updates after each initial interview. The last 15 interviews used the final version of the interview guide, which version is provided in Appendix~\ref{sec:guide}. The interview guide consists of five parts: interviewee information, case context, CE process, experiment details, and holistic CE view. The guide was designed with probes such that it could be adapted to the interviewee's background and role.

We \emph{sampled and selected} case companies through a mix of convenience sampling and snowballing guided by the aim of our study. We selected companies that either apply CE or have plans to start applying CE. In three of the cases we selected a branch of the company as the case (cases D, I and G). Candidate companies were identified through searching on LinkedIn and Google for job listings for data scientists where A/B testing was mentioned, through personal contacts, and through asking the interviewees (i.e. snowballing) for other companies involved with CE. As such, the cases were not restricted by geographical location. Our aim was to interview practitioners with insights and experience of applying CE, and primarily in the roles with heavy CE involvement: software developers, product owners, and data scientist. We applied snowball sampling within the companies by asking for additional interviewees within the companies during the initial interviews.  The managers closest to CE were contacted via e-mail with information about the study and a request to perform interviews. In total 35 companies were contacted of which 12 were included in our study. Of the remaining, 11 did not respond and 13 were not applicable. A description of each included case company is provided in Section~\ref{sec:method_cases}.

The \emph{semi-structured interviews} were held during an initial 3 month period for the pilot study, which was then continued a year and a half later for another year. Interviewees were selected from the case organisation until a complete picture of how and why they used CE was obtained. Albeit in some cases (B and E) the process was cut short due to the subsequent interview prospects not wanting to be interviewed. Each interview was approximately 60-120 minutes long and was held as an open conversation in-line with the interview guide. The interviews were audio recorded after permission for this was granted by the interviewees. The majority of the interviews were performed at the companies' premises. For six companies, this was not feasible due to their location, in which case the interview were held on-line via Zoom or Skype. After the interviews, the audio recordings were transcribed word-by-word into 231 pages. Some of the interviews were done in Swedish and quotes from those interviews were translated to English.

\emph{Thematic coding} was performed on the interview transcripts using the pre-defined codes of our code book. The code book was defined based on previous knowledge and insights into CE, from primarily three core publications~\cite{bosch2012building,fagerholm2017right,kohavi2009controlled}. It was iteratively refined throughout the coding process by discussing these codes within the group of authors. The codes were added per paragraph of transcribed text, and each paragraph could have multiple codes. The codes were clustered into 11 themes according to thematic analysis. The codes cover areas such as the company context, management of data, the business model, product, and CE process for the company, see Appendix~\ref{sec:codes}. After having coded the full set of transcripts, cross-case analysis was performed to identify factors and patterns common to several cases. The theory was gradually inducted as this cross-case analysis matured, see below.

\subsection{Theory building}
\label{sec:method_theory_building}
We have built the theory by analysing our empirical data and gradually defining constructs, propositions, the scope and explanations for the theory, based on guidelines provided by Sj{\o}berg et al.~\cite{sjoberg2008building}. The final step of theory building, namely testing our theory, remains as future work. In this article, we provide an initial validation of our theory by characterizing each of our case companies based on the theory. In this way, we illustrate that the theory is useful for categorising and explaining an organization's ability to perform CE.The first author defined the initial version of the theory through analysis of the empirical data. The theory was then refined by the first and the second author through multiple iterations of review and discussions. As the theory matured, it was reviewed by all three authors and further improved.

The \emph{constructs and propositions} of FACE represent the factors that affect CE and the relationships between these factors, relevant to CE. The constructs were identified through analysis of the thematically coded interview data. The themes from thematic analysis formed the first iteration of the constructs of the theory. An orthogonal coding step was carried out afterwards to find relations between constructs, which formed the propositions; using each theme-pair as a code. The constructs and the propositions were then gradually refined and adjusted through discussion within the team of authors, to provide a clear and concise description of the factors that affect a company's ability to draw benefits from CE. Describing the relationships between the constructs through visualisation (see Figure \ref{fig:theory}) and written definitions (see Sections~\ref{sec:theory_constructs} and \ref{sec:theory_propositions}) facilitated this inter-author discussion. The supporting evidence in the material was used as a basis for the discussion. In total, 10 iterations of the theory were discussed in this way.

The \emph{scope} and \emph{explanations} of FACE were identified based on our empirical data. The \emph{scope} of our theory was defined through considering the common characteristics of our case companies, and thus the type of organizations that our theory may be applicable to. Explanations for our theory are provided as part of the definition and description of each construct and proposition.

The \emph{empirical underpinning} of FACE provides a motivation for the concepts of our theory grounded in the empirical data, thereby illustrating the empirical foundation for the constructs and propositions. In Section~\ref{sec:results}, supporting evidence in the material for each proposition in the theory is laid out along with expanded explanations. The empirical underpinning in terms of the constructs and propositions provides an \emph{initial validation} of FACE, and illustrates its utility and explanatory power. As such, the empirical data is used both as a source and validation, by having the data used at the detailed level to construct the theory and then at the holistic level to describe our set of companies.


\subsection{Case companies}
\label{sec:method_cases}

The twelve case companies in the study differ on many attributes such as size, product domain, business model and CE practices. Our case companies range from small to huge multi-national companies. Some of the companies work extensively with CE while some do next to no experiments at all. 

Table~\ref{tab:companies} contains an overview of the business models and states of CE practices for the 12 case companies. Each case company is ranked according the its \emph{expertise} with CE and the \emph{extent} to which the company conducts experiments (both frequency of performing experiments and on what parts of a product or service that are experimented on). The business model is given as a brief summary for each company, primarily describing the company's offering and what customers that are targetted. Under the business model column, \textit{direct sales} refers to having a business model where licenses to the product or service are sold by contacting customers directly.

The 27 interviewees at the case companies are described in Table~\ref{tab:interviewees}, which show their code corresponding to the case company and role name. Two additional roles have been observed in the interviews in addition to the ones presented in RIGHT~\cite{fagerholm2017right} (see Section~\ref{sec:background_ce}), namely, the \emph{User Researcher} and \emph{UX Designer}, These roles have similar responsibilities within CE to that of data scientists and software engineers, respectively. That is, a UX designer or software engineer comes up with a change in user experience or software, and a user researcher or data scientist analyzes the results. The user researchers in the interviews primarily used prototyping qualitative experiments. The actual titles differ significantly from their assigned role and the titles in use include, e.g., head of growth, software engineer, head of customer success, growth engineer, head of research, etc.

\begin{table}[t]
    \sf
    \caption{Overview of our 12 case companies, containing business model, size and age (rounded to nearest 5 year), and state of CE. Small companies have less than 50 employees, medium have less than 250, large companies have less than 1000 employees, and huge more than 1000 employees. CE expertise and extent is estimated based on our interviews using the ordinal scale: low, medium, and high.}
     \label{tab:companies}
   \centering
       \renewcommand{\arraystretch}{1.2}

    \begin{tabularx}{\textwidth}{@{}l@{~}p{1.7cm}X>{\centering\arraybackslash}lcccc@{}}
    \toprule
        &&  \multicolumn{2}{c}{\textbf{Business Model}} &&& \multicolumn{2}{c}{\textbf{CE}} \\
        \cmidrule{3-4}\cmidrule{7-8}
        \multicolumn{2}{@{}l}{\textbf{Case}} & Summary & Type & \textbf{Size} & \textbf{Age} & Expertise & Extent \\
        \midrule
        A & E-commerce algorithms &  Product with direct sales & B2B & Small & 20 & High & Medium \\ 
        B & Local search & Service selling placement to B2B customers and serving ads to B2C & B2X & Small & 15 & Medium & Low \\ 
        C & E-commerce consultants & Consulting on a per project basis & B2B & Medium & 20 & Low & Low  \\ 
        D & Video streaming &  Freemium service with premium features & B2C & Medium & 10 & High & Full \\ 
        E & Web shop & Web shop for subscription to physical goods & B2C & Huge & 10 & Medium & Medium \\ 
        F & Customer relations & Product and service with direct sales & B2B & Small & 30 & Low & Low \\ 
        G & Engineering tools & Freemium service with premium for larger teams & B2B & Huge & 20 & High & Full \\ 
        H & Web shop & Retailer with small web shop & B2B & Large & 20 & Low & Low \\ 
        I & Web shop & Web shop and retail & B2C & Huge & 20 & High & Medium \\ 
        J & \makecell[tl]{Product \\ information} & Service selling customer information and leads to B2B and free for B2C & B2B & Medium & 10 & Medium & Low \\ 
        K & \makecell[tl]{Business \\ intelligence} & Product with direct sales & B2X & Large & 30 & Medium & Low \\ 
        L & \makecell[tl]{Employee \\ management}  & Service with direct sales & B2B & Small & 20 & Low & Medium \\ 
        \bottomrule
    \end{tabularx}
\end{table}

\begin{table}[t]
    \centering
    \sf
    \renewcommand{\arraystretch}{1.2}
    \caption{Overview of the 27 interviewees at the case companies. The codes of the interviewees correspond to the case organisation that they belong to. The role is assigned by the authors to best describe their primary role(s) in CE---not their title.}
    \label{tab:interviewees}
\begin{tabularx}{\textwidth}{@{} r@{\hskip 2mm}>{\raggedright\arraybackslash}X @{}l@{\hskip 4mm}l@{} r@{\hskip 2mm}>{\raggedright\arraybackslash}X @{}l@{\hskip 4mm}l@{} r@{\hskip 2mm}>{\raggedright\arraybackslash}X@{}}
    \toprule
    \textbf{Code} & \textbf{Role(s)} &&& \textbf{Code} & \textbf{Role(s)} &&& \textbf{Code} & \textbf{Role(s)}\\
        \cmidrule{1-2}\cmidrule{5-6}\cmidrule{9-10}
        A1 & Software Developer and Data Scientist &&& D2 & User Researcher and Software Developer &&& I2 & Product Owner \\
        A2 & Product Owner and Data Scientist &&& E1 & Data Scientist and  Software Developer &&& I3 & Data Scientist \\
        A3 & Software Developer and Release Engineer &&& F1 & Product Owner and Business Analyst &&& I4 & Business Analyst \\
        A4 & Product Owner and Business Analyst &&& F2 & Product Owner and  Software Developer &&& J1 & Product Owner \\
        A5 & Operations Engineer &&&  G1 & Quality Assurance &&& J2 & Software Developer and Data Scientist \\
        B1 & Software Developer and Data Scientist &&& G2 & Product Owner &&& K1 & Product Owner and UX~Designer\\
        C1 & Software Developer &&& G3 & Data Scientist &&& K2 & User Researcher \\
        C2 & Software Developer and Quality Assurance &&& H1 & Software Developer &&& L1 & UX~Designer \\
        D1 & Data Scientist and Business Analyst &&& I1 & Software Developer &&& L2 & User Researcher \\
        \bottomrule
    \end{tabularx}
\end{table}

\subsubsection{Case A: E-commerce algorithms}
\label{sec:case_a}
This case company offers an e-commerce platform that is sold to companies (B2B). The platform consists of various algorithms for ranking products and an administration interface. While the algorithms target end-consumers, the administration interface targets managers at the e-commerce companies. The algorithms provide value for the end-consumers by increasing the relevance of the product that they see on the web shops. The case company is fairly small with about 50 employees and was established 20 years ago. The company's business model is to sell usage licenses to other companies and the company has salespersons working with direct sales as their only source of revenue. Company A has experienced several periods of growth after which the company has had to scale down due to failure of a big sale. The company conducts a medium amount of CE but only on the software that targets their end-user consumers, not on the administrative interface. The company only performs quantitative experiments since the software that is experimented on does not have a graphical interface. Company A also assist their business customers with their CE.

\subsubsection{Case B: Local search service}
\label{sec:case_b}
Case B offers a search engine service for search within a local region, a.k.a., yellow pages. The search engine is free to use and the major source of revenue is companies buying promotion in the search result rankings and visual presentation. The case company has a large sales team that work with direct sales by calling potential business customers. The company is about 20 years old and has stayed stable for some years at about 50 employees. The company was recently acquired by a larger business group and additional products were integrated with the local search service, such as a ticket booking service. CE is not frequently applied at the case company and is only used to verify large changes to the search engine ranking algorithm with quantitative data.

\subsubsection{Case C: E-commerce consultants}
\label{sec:case_c}
Case company C is a consultancy firm that develops and maintains web shops for other companies with relatively large demands on traffic volume and/or product catalogue. The work includes a lot of integrating various software systems, e.g., product information management (PIM), content management systems (CMS), payment gateway systems, search or recommendation engines, etc. These software systems each come with their own administrative tools and the case company helps their customers to use these tools. Company C has built its own software to optimize and support the process of building the web shops, but the business model is to sell consulting hours in a per project basis. The company has existed for more than 20 years, currently has about 100 employees and is experiencing steady growth in the number of employees. The company does not conduct any CE on its own software but has assisted its customers in conducting quantitative experiments on their web shops.

\subsubsection{Case D: Video sharing}
\label{sec:case_d}
Case company D develops and sells a video sharing platform where users can record and edit videos for marketing purposes. The company has been operating for 10 years, and has about 200 employees of which about 30 people are in the software development department, distributed over four teams. Recently, the company pivoted its business model by adding a new product that is targeted at individuals; aimed at consumers and smaller companies that wish to market themselves. This new product is the defining boundary of case D. Prior to the pivot, the customers have mainly been other businesses (B2B) that Company D has reached through a fairly big sales department with direct sales.  The new product has an entirely separate development department and no sales persons are involved. The pivot enabled the use of CE at the company. The new B2C product is offered under a freemium license, where the free version is offered with limited video uploading capacity, and the paid version offers additional features. The B2B part of the company performs no CE. In contrast, the B2C team conducts extensive CE and has various specialized roles for supporting CE and related activities, viz., data scientists and data engineers for quantitative experiments and user researchers for qualitative experiments.

\subsubsection{Case E: Web shop}
\label{sec:case_e}
This case company offers a web shop with a subscription service to physical goods. The web shop drives sales through their online presence only and does not have any retail stores. The company was founded less than 10 years ago and has experienced rapid growth in revenue and number of employees. At the time of the interview, the copmany has around 3\,000 employees, most of whom work with delivering the physical goods. The company's IT department accounted for about 250 employees. In addition to the web shop, the IT department also operates inhouse software systems for handling logistics, marketing, etc. The interviewee at the company was involved with optimizing multiple of these products at the company. There is a drive from the management to be data-driven and CE is encouraged on a strategic level. However, this is hindered in practice by chaotic management, caused by the rapid expansion.

\subsubsection{Case F: Customer relations product}
\label{sec:case_f}
The next case sells a customer relations product (and service) to other businesses. The product is highly customizable and each new customer gives rise to a new integration project. The integration includes adapting to another company's data model and internal software systems, e.g., for human resources. The company is 30 years old and has 200 employees, 35 of which are in software development, divided into three teams. Sales is a large part of the company, and all sales are done through direct sales. Half of the company is committed to pre-sales and operations that support customers before and after sales due to the complex integration. No free version or trial of the product is available. 
Compny F has conducted a few prototyping qualitative experiments on recent new features, but not on all developments. The company is aware of the concept of quantitative experiments but has not conducted any and have no concrete plans for applying CE in the near future. However, the company has some of the technological pre-requisites to conduct post-deployment CE in place already, and make use of feature flags to verify new developments at specific customers, but not in a systematic way.

\subsubsection{Case G: Software engineering tools}
\label{sec:case_g}
Case G is a huge international company with multiple products that focus on supporting the software engineering process. The main products are a project tracker, issue tracker, and a team collaboration platform. The company has existed for roughly two decades and today have more than 4\,000 employees. The majority of these employees are within IT. Each of the company's products has its own development organization. The company advocates agile software development. Notably the company has no large sales department and do not use direct sales at all, despite their focus on business to business (B2B). The company conducts huge amounts of experiments on all aspects of their products---both on new and old products, and makes use of both quantitative and qualitative experiments. A specialised team supports product teams in applying a CE approach. Since there are multiple teams involved with CE on multiple products, the company has an extensive CE platform. The company also has a formal process for conducting CE developed by the CE team.

\subsubsection{Case H: Web shop}
\label{sec:case_h}
This case company offers a web shop for business customers only. The company has over 2\,000 employees, of which most are employed at retail stores. The company used case company C to develop their web shop and also has their own small IT department that manages and improves the web shop. The company has conducted ad hoc experiments on occasion and uses off-the-shelf CE tools (Google Analytics) for supporting their CE. The biggest limiting factor of the Company H's CE practices is the lack of human resources at the IT department.

\subsubsection{Case I: Web shop}
\label{sec:case_i}
The next case is a huge international conglomerate. The company has a long and rich history in retail and have operated a web shop for over 20 years. The web shop is becoming increasingly important to the company. The company has about 200\,000 employees of which about 5\,000 are employed in the IT organization. The company have many products for managing their logistics, etc., but in this study we focus on their web shop product only. Several cross-functional development teams are responsible for various parts of the web shop (such as the recommender engine). Not all parts of the web shop are experimented on, but the trend is moving towards more CE work. There are multiple teams involved with CE at the company, some of them have their specialized CE infrastructure but most teams use a centralized experimentation platform and contact a data science team for help with CE.

\subsubsection{Case J: Product information platform}
\label{sec:case_j}
Case J offers a service for product information within the building industry that is free to use. The company was established less than 10 years ago. The source of revenue come from customers that want to know who accessed their product's information in order to obtain sales leads. The case company describes their business model as being a middle man that sells information, and has a sales department that works with direct sales. The company also relies on organic growth based on their free users. There are about 200 employees at the case company, half of which work within the IT organization. Company J does not make use any of qualitative methods and quantitative experiments are new to the company. Currently, the company does not experiment on all parts of their software. There is a small and dedicated team with two employees that conducts experiments, and the company expects that the other software engineering teams will soon start applying CE. 

\subsubsection{Case K: Business intelligence}
\label{sec:case_k}
This company offers several advanced business intelligence products for different needs. The products are used to make graphs, tables, and other visualizations from various data sources. The products are advanced to use and the flagship product even has a proprietary domain-specific programming language for data manipulation. Since the company is more than 20 years old, the products are in different stages of the life cycle, albeit all of them are still offered. The company is a large enterprise with about 3\,000 employees and 500 of them in the development department. The company has grown rapidly during the latest years with an increase in the number of employees. The sales department is very large and uses primarily direct sales to other businesses. The company has a team specialized in user experience research that gathers qualitative feedback on a regular basis. The company conducts qualitative experiments to evaluate both prototypes and completed functionality with users. However, not all development teams are on board with this yet so not all features are evaluated in this way. The user experience research team is aware of and interested in quantitative experiments but getting such systems in place has not been a company priority.

\subsubsection{Case L: Employee management product}
\label{sec:case_l}
The final case company develops a product for employee management. The product is intended to be embedded in the customers' intranet and has both administrative and end-users within each customer organization. The company was founded 20 years ago and has about 250 employees of which about a third in the IT department. The company relies on direct sales with a sales force and most of the revenue comes from projects that integrate their product at customer sites. The company pays close attention to user experience and has a team involved with user research that conducts qualitative experiments with prototypes. Since the product is still considered to be early in development this CE is somewhat frequent.

\section{Theory Formulation of FACE}
\label{sec:theory}

\begin{figure}[htbp]
    \centering\sf\footnotesize
    \input{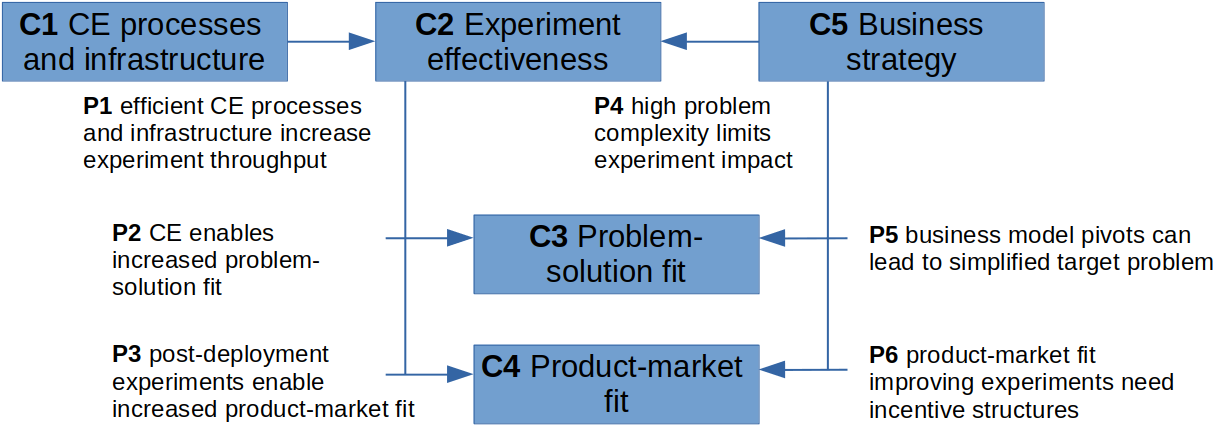}
    \caption{The constructs and propositions of the Factors Affecting Continuous Experimentation (FACE) theory. The boxes represent constructs and the arrows are propositions.}
    \label{fig:theory}
\end{figure}

Our FACE theory describes factors that affect continuous experimentation (CE) and how these factors contribute to an organization's ability to gain value through CE. The gains are achieved through effectively conducting experiments that enable improving the problem--solution fit (\textbf{P2}) and/or the product--market fit (\textbf{P3}). In essence, the theory states that the effectiveness of experiments can be increased through efficient processes and tools for CE (\textbf{P1}), addressing a user problem that is sufficiently simple to be measurable (\textbf{P4}), pivoting the business model to simplifying said problem complexity (\textbf{P5}), and/or selecting a business model that provides incentives to conduct experimentation (\textbf{P6}). An overview of our theory is provided in Figure~\ref{fig:theory}. In this section, the theory is defined by describing its constructs, propositions,  scope, and its validity is discussed. Expanded and empirically-based explanations of the theory are provided in the succeeding Section~\ref{sec:results} by exemplifying the constructs and propositions as observed in our 12 case companies.

\subsection{Definitions}
\label{sec:theory_definitions}

The following terms are used to describe the constructs and propositions of FACE theory:
\begin{description}
\item[\textbf{Experiment}] is an activity that introduces a \emph{change} in a product or service \emph{business offering} with the \emph{goal} of learning or improving the offering based on feedback through \emph{user data}. The term covers various types of experiments such as quantitative controlled experiments (A/B tests) and quasi-experiments, and includes less rigorous prototype experiments where the users' ability to use a specific feature or willingness to pay for a feature is evaluated qualitatively. There must always be a way to decide if the \emph{goal} of the experiment is met or not, otherwise it is not an experiment by our definition. In the case of a controlled experiment, this goal is quantitative and assessed using metrics derived from \emph{user data}.
\item[\textbf{Qualitative experiments}] can be done by implementing a simpler prototype and evaluating it on users. The fidelity of the prototype can be anything from a hand-drawn sketch to almost completed functionality. It is hard to get sufficient numbers of users to evaluate a prototype quantitatively in a controlled experiments, so pre\=/deployment experiments are usually evaluated with qualitative data.
\item[\textbf{Quantitative experiments}] are executed live in a production environment with real users. As such, quantitative experiments usually have higher infrastructure needs than \emph{qualitative experiments}.
\item[\textbf{Continuous experimentation (CE)}] is the process of continuously using experiments. This process encompasses the whole software engineering process and involves multiple experiments conducted in iterative cycles. Based on observations from the interview material, we deduce that experiments are described as \emph{continuous} for two reasons. First, an initial \emph{experiment} is not always decisive, and many experiments in a row might be needed to refine a \emph{change}. Second, the experiment result might uncover new knowledge about the business offering, market, or users that begets further related inquiries or even new features.
\item[\textbf{Experimenter}] is a person initiating and or being in charge of an experiment. It is not a formal role in the RIGHT model (see Section~\ref{sec:background_ce}), but would usually correspond to the data scientist or business analyst.
\item[\textbf{Change}] refers to a modification to a \textit{business offering}. It can be a new functionality in the software (i.e. a \emph{feature}) or a modification to an existing one, or a change to the quality of the software. The change can be in any part of the software and they can be subtle for users to perceive. The scope of the changes we have observed in the interview material varies, from a small tweak of the font-size on a button to rebuilding a large part of the product\footnote{Many authors caution against large changes in an experiment to minimize risks of errors~\cite{fagerholm2017right,kohavi2009controlled}.}. For some companies a change corresponds to a commit in a version control system~\cite{kevic2017characterizing}.  A change corresponds to a \emph{treatment} in traditional statistics or medicine literature. 
\item[\textbf{Goal}] is what an experimenter wants to achieve with an experiment. In a quantitative experiment, the goal can be described with a measurable improvement to a \emph{user experience} or \emph{sales metric}. The term \emph{goal} is preferred in this work, since the more specific term \emph{hypothesis} implies a deeper thought behind an experiment, which is not always the case. For example, when Google tested 41 shades of blue for their links~\cite{holson2009putting}, the goal was to find the colour that got the most clicks, but the learning obtained is limited and the hypothesis is unclear.
\item[\textbf{Business offering}] is a catch-all phrase referring to either a product, service, web shop, etc.  This term is used when the distinction between these are not important.
\item[\textbf{User data}] refers to data obtained from actual users of the business offering. User data could also be obtained through other means, such as questionnaires, interviews, or by eye-tracking. There are two categories of metrics derived from user data: (1)~\textit{user experience metrics} and (2)~\textit{sales metrics}.
\item[\textbf{User experience metrics}] are derived from \textit{user data} that measure the users' experience, such as degree of users that engage with the software during the experiment, time spent on parts of the software, rate of users that complete a certain task etc.
\item[\textbf{Sales metrics}] are derived from user data originating from the sales process of the offering. The sales metrics include revenue from sales or subscriptions, churn of subscribed users, conversion rate of users to paying users, etc.
\item[\textbf{Proxy metric}] is a metric that substitutes for another more relevant metric that cannot be used directly in an experiment for some reason. For example, consider an e-commerce web shop with insufficient traffic to obtain statistical significance on conversion rate or other sales figures. They could use clicks (i.e. a user experience metric) as a proxy metric for purchases (i.e. a sales metric) since far more users click on products than buy them. However, the signal-to-noise ratio would be much lower and an increase in clicks could even lead to a decrease in purchases if the user experience becomes more complicated as a result of the change (requiring more clicks). As such, the assumptions made when choosing a proxy metric should be continuously verified.
\item[\textbf{Business model}] is the way a company has structured their activities, revenue streams, costs, etc., to obtain and give value to users, see Section~\ref{sec:background_business_models}.
\item[\textbf{Pivot}] is a major change in the business model done in order to better realize the business strategy~\cite{chaparro2021pivot,kirtley2020pivot}. This may involve \emph{changes} to the product offering (which is part of the business model), but a pivot could also entail changes to, e.g. revenue model or sales channels. The distinction we draw between \emph{changes} and \emph{pivots} is that pivots have a larger scope and are often not based on evidence to the same degree as a change involved in an experiment. A pivot can be done for various reasons, such as to find a different user base or to overhaul the business offering in a major way.
\end{description}

\subsection{Constructs}
\label{sec:theory_constructs}

FACE identifies that both CE processes and infrastructure (\textbf{C1}) and the complexity of the problem that the offering solves for users (\textbf{C5}) have an effect on an organization's ability to experiment effectively (\textbf{C2}), in a way that affects the problem-solution (\textbf{C3}) and product-market fit (\textbf{C4}). Furthermore, FACE states that an organization's incentive structures (\textbf{C6}) play an important role in facilitating experimentation directly through providing a link between users and organization.

\begin{description}[listparindent=\parindent]
\item[\textbf{C1}] \emph{CE processes and infrastructure} are software engineering practices enabling \textit{continuous experimentation}, primarily through providing technical infrastructure and processes for conducting experiments, but also through organizational infrastructure that provides necessary competences and roles. Since continuous experimentation involves implementing changes in the software for a business offering, this construct also encompasses standard practices for efficient software engineering, such as software testing and quality assurance, while it is not the focus of this study.

The infrastructure includes a continuous integration and continuous deployment (CICD) pipeline that enables experiments to be executed in an efficient manner, infrastructure to store and extract data, and an experiment platform to manage and orchestrate experiments running in parallel. The organizational infrastructure involves having access to all required CE roles. CE processes include processes for designing and conducting experiments, and for decision-making where experiments are integrated in the overall context of a company. For example, they prescribe how hand-over is handled between experiments initiated by data scientists or marketers to the software development department. Conducting experiments in an effective manner requires a degree of rigor in the processes, such that experiment results are accurate and trustworthy. The processes do not have to be explicitly documented to be rigorous, as long as they are followed.

\item[\textbf{C2}] \emph{Experimentation effectiveness} is the degree of effective continuous experimentation conducted in an organization. It is a function of the throughput of experiments, the impact the experiments have on users or business, and the ability to accurately measure this experiment impact. Throughput is determined by experiment speed and experiment capacity, where speed is the time it takes to perform an experiment from designing it to making decisions based on its result. Experiment capacity is how many experiments that an organisation can handle simultaneously. The throughput of experiments is affected by both the organizations' capabilities and the amount and type of user data available. 

\item[\textbf{C3}] \emph{Problem--solution fit} is the degree to which an offering provides a solution to users' problems. Problem and solution are part of the lean canvas business model~\cite{maurya2012running,ries2011lean}. There can be multiple problem--solution pairs in the same offering and a solution can address multiple problems. The degree to which the problem--solution fit can be measured depends on the specific offering but is usually measured with  \emph{proxy metrics} based on \emph{user experience metrics}.

\item[\textbf{C4}] \emph{Product--market fit} is ``\emph{[a] measure of how well a product satisfies the market.''}~\cite[p.~7]{olsen2015lean}. This construct describes the ability of an offering generate economic value (usually revenue from sales) and thus that there is a market for the offering. Product--market fit is advocated in lean startup as the ultimate target for business success for all companies~\cite{maurya2012running}, not just startups. It may be hard to directly measure product--market fit in an experiment through instrumentation of the offering, but it can be obtained through customer surveys~\cite{ries2011lean}. Sales figures in the form of user retention, growth of the user base, etc., can serve as \textit{proxy metrics} for product--market fit.

\item[\textbf{C5}] \emph{User problem} is what problem the users solve with the \emph{business offering} and how the problem is solved.

\item[\textbf{C6}] \emph{Incentive structures} is the way an organization has structured rewards (or punishments) related to the business model and the value proposition therein. The construct primarily concerns the degree to which users are enticed to pay for, and employees are enticed to improve, the \emph{business offering}. This includes, for example, the licensing and revenue model under which the software is sold and how the performance of software developer teams are measured.

\end{description}

\subsection{Propositions}
\label{sec:theory_propositions}

The propositions describe how the constructs influence an organization's ability to perform effective experiments (\textbf{P1} and \textbf{P2}) that in turn indirectly (\textbf{P3}) or directly (\textbf{P4}) affects the product--market fit of their business model through continuous experimentation (\textbf{P5}).

\begin{description}[listparindent=\parindent]

\item[\textbf{P1}: \textbf{C1}$\rightarrow$\textbf{C2}] \emph{Efficient CE processes and infrastructure increase experimentation throughput.} The existence of processes and infrastructure (\textbf{C1}) that support CE increases an organization's experiment effectiveness (\textbf{C2}). For example, a continuous experimentation platform could enable performing parallel experiments, thus increasing the number of experiments, or an efficient continuous integration and continuous deployment pipeline could ensure that deployments happen without delay, thus increasing the speed of experiments. Additionally, the degree to which an organization has prepared their data infrastructure for collecting and analyzing data in their software offering and sales process determine the extent to which they can experiment on those parts.

As for processes, correct prioritization of experiments ensures that the changes with the highest potential impact on problem--solution fit (\textbf{C3}) or product--market fit (\textbf{C4}) are selected, that the experimenter and the organization can trust the reliability of the experiment, that the metrics used are relevant, etc. Continuous experimentation is an activity that encompasses most aspects of software engineering, from prioritization of features to post-deployment, and any delay to any of those activities will cause delays also in continuous experimentation too. Hence, there are many opportunities for improving the effectiveness of experiments, e.g. experiment throughput, (\textbf{C2}) through processes and tools or other infrastructure for continuous experimentation.

\item[\textbf{P2}: \textbf{C2}$\rightarrow$\textbf{C3}] \emph{CE experiments enables increased problem--solution fit.}
Experiments (\textbf{C2}) that target problem--solution fit (\textbf{C3}) aim to improve the users' ability to solve problems with the product, and thus also improve the user experience. Either \emph{qualitative experiments} or \emph{quantitative experiments} can be used for this purpose.

The process of conducting effective experiments (\textbf{C2}) over time to optimize the product towards a defined goal will lead to improvements in the product based on that goal. This is the main benefit of CE. The impact of each individual change might not be great. However, since the changes introduced in an experiment are only finalized in the offering if they have a positive impact, then the accumulated experiments conducted over time will have a positive effect. There are diminishing business value returns to how well the problem--solution fit can be achieved. Even if the solution perfectly solves the users' problem, it does not mean that the problem is important and that users are willing to pay for it.

\item[\textbf{P3}: \textbf{C2}$\rightarrow$\textbf{P4}] \emph{Quantitative experiments enables increased product--market fit.}
The product--market fit (\textbf{C4}) measures how well the product can meet market needs and generate revenue, hence it is the ultimate goal of for-profit products or services. The companies that are able to affect product--market fit can use CE to great advantage. However, product--market fit is more difficult to target than problem--solution fit due to having to affect customers' overall purchasing intent. The changes usually only have a subtle effect on product--market fit and so require precise measurements that only quantitative data can serve, i.e., with effective controlled experiments (\textbf{C2}).
Not all companies are able to affect product--market fit with experiments due to either high user problem complexity (see \textbf{P4}) or lacking incentive structures (\textbf{P6}).

\item[\textbf{P4}: \textbf{C5}$\rightarrow$\textbf{C2}.] \emph{High problem complexity limits experimentation impact. } 
To be effective, the change being experimented with should have a direct link to the actual user or business value, thereby enabling assessment of the impact of the experiment (\textbf{C2}). Establishing that link can be harder for \textit{business offerings} solving complex user problems (\textbf{C5}), such as software in a car, rather than for simpler problems, such as a webshop. This is due to it being harder to modify the product and to accurately measure the impact of a change for complex user problems. Thus, in order to perform experiments under those circumstances, the possible changes are limited or the impact of the changes will be uncertain, respectively. Post-deployment experiments are more challenging and are thus affected by high problem complexity to a higher degree.

Reasons for the product being hard to modify could be due to not wanting to disrupt users because they need stability, rigid customer requirements and contracts, complex deployment and delivery with on-premise installations to other organizations, or lack of access to software that the product is integrated with. The product can be hard to measure if the usage is hard to quantify, the product is configurable with variants that the proposed change affects differently, or there is no meaningful metric to use. These examples are all aspects of problem complexity of the problem-solution pair from the business model, either directly or indirectly through the ways that the product is built to deal with problem complexity (i.e. configurability).

\item[\textbf{P5}: \textbf{C6}$\rightarrow$\textbf{C5}.] \emph{Pivots can lead to simplified target problem}
CE is suitable for changes that address the solution part of the problem--solution pair (\textbf{C3}). However, making changes to address the problem is a big endeavour because is is a big change and will likely require involvement of the whole company (i.e. software engineering, sales \& marketing, and management). CE is not suited for such large changes because the proposed change are too costly to reverse in case of failure. In addition, according to \textbf{P4}, the presence of high problem complexity itself limits the effectiveness of experiments. Consequently, experiments might not be useful for changes that affect the problem complexity. Instead, a \emph{pivot}, i.e., a change of direction in the business model that affects the value proposition and thus the product can be required to realize changes to user problem complexity (\textbf{C5}) and hence the ability affect problem--solution fit (\textbf{C3}) with CE.

An example of a pivot would be to shift from targeting a broad market to addressing a narrower customer segment consisting of a niche market with a more specific value proposition. This would hopefully reduce the set of required features and thereby reduce the problem complexity and the user experience could be simpler.

\item[\textbf{P6}: \textbf{C6}$\rightarrow$\textbf{C4}] \emph{Product-market fit improving experiments need incentive structures.}
Problem\--solution fit (\textbf{C3}) can be increased by finding user problems and designing solutions for them. Software engineers used to an agile user\-/centered process will be familiar with this approach~\cite{brhel2015exploring}. However, targeting product--market fit (\textbf{C4}) is more challenging, since market considerations are outside traditional software engineering responsibilities. Also, sales metrics might not be available for CE depending on the licensing and revenue model. As such, targeting product-market fit requires incentive structures to be in place with a link between business value, user value, and software engineering activities. This can, for example, take the form of an explicit team goal on a target metric that can affect product\--market fit and be measured in experiments, thereby persuading the team to take action outside their traditional responsibilities.

Note that a focus on product--market fit does not come with a degradation in user experience or the users' obtained value of the software. Indeed, if there is a direct connection between the users' satisfaction with the software and their inclination to pay for it, then sales figures can be used to simultaneously optimize user \emph{and} business value. This is the case for software provided as freemium  (where the user can use a free or a paid version of the software), as a demo/trial versions, as a subscription, with in-app purchases, etc. This is not the case for business models where the user pays once upfront for the product without first using it.

The following consequences on product--market fit experiments happen when the incentive structures are in place:
\begin{itemize}
    \item The software engineering  and sales \& marketing departments will have the same incentives and can thus work more easily together. This is also evident from the rise of growth marketing/engineering in industry~\cite{kemell2019software,troisi2020growth}.
    \item The prioritization process can use metrics from the sales process to inform development and can chose to develop features they believe will best benefit users and business.
    \item Focusing on acquiring new users will lead to having more user data which will increase the ability to do experiments further.
\end{itemize}

\end{description}

\subsection{Scope}
        
The scope within which FACE is applicable is user-intensive software companies with a user-facing offering. For example, media content services (Case D), e-commerce web shops (Cases E, H, and I), and application software products (Case K). Not all companies in the study have software development as their primary activity, but software is a central part of their business model. The theory is derived from empirical observations on those companies and it is unknown whether the theory is applicable or useful outside this context of software intensive companies, for example, non-profit organizations developing open-source software, companies developing embedded systems, companies that are involved with CE from a marketing perspective but that do very little software development, such as a news website or web shops without IT departments.

\subsection{Theory validity}
We evaluate FACE using the criteria proposed by Sj{\o}berg et al.~\cite{sjoberg2008building}, namely testability, empirical support, explanatory power, parsimony, generality, and utility. 

\begin{description}
\item[Testability.]
FACE makes high level claims about company processes, organization, etc. As such, the ability to use experiments to test the theory is low due to the large scope that would be required of such an experiment. However, the theory makes claims about real world phenomena and those claims can be validated by using it to analyze and explain CE practice at other companies.

\item[Empirical support.]
The theory is based on an extensive multi-case study with 12 cases of various contexts and 27 interviewees with various backgrounds and roles. The cross-section of empirical underpinning per case and proposition is shown in Table~\ref{tab:empirical_underpinning}. Each proposition is supported by evidence from multiple cases.

\item[Explanatory power.]
While there is no ability to make quantitative predictions from FACE, the theory can be used to differentiate CE in real companies and can explain why some companies derive more value from CE than others. The context of all case companies that the theory is based on is also available (see Section~\ref{sec:method_cases}) and can be used by practitioners to compare with their own organization's context, and thus judge the applicability of the theory for that context using theoretical generalisation. The theory also uses terms and concepts from established pre-validated knowledge, such as lean canvas from lean-startup~\cite{maurya2012running,ries2011lean}.

\item[Parsimony.]
The number of constructs and propositions have been continuously reduced and combined during the theory building process. The remaining constituents of FACE are needed to explain the data from the cases.

\item[Generality.]
The scope of the theory is user-intensive companies which limits the generality of the theory to such companies. The case companies on which the theory is based are quite different in terms of size, age, and domains (though e-commerce is over-represented) so the breadth of the scope is wide.

\item[Utility.]
FACE can be used by software organizations to understand their ability to perform CE and the factors that influence this. The interviewees in the study were generally very interested in learning more about CE which hints at the overall industry relevance, in addition to the large number of industry authors active in the research about CE~\cite{auer2018current,ros2018continuous}. 

\end{description}
\section{Theory Explanations and Empirical Underpinning}
\label{sec:results}

In this section, the supporting evidence from each of the 12 cases, for each of the six proposition in FACE is presented in order, along with an expanded explanation of the meaning and impact of each proposition. We refer to theory constructs as \textbf{C1}--\textbf{C6} and propositions as \textbf{P1}--\textbf{P6}, and \textit{italicize} the key terms of our theory when mentioned in the text. Note that, only the salient evidence is discussed in the text. See Table~\ref{tab:empirical_underpinning} for a complete picture of the supporting evidence per case and proposition. The strength of the evidence is judged qualitatively based on how much and how direct the propositions are discussed at the interviews. As such, it does not reflect, e.g., how good or bad the cases are at conducting experiments (\textbf{P1}).

\begin{table}[t]
    \sf
    \caption{\setlength{\fboxsep}{1pt}Cross-section of empirical underpinning per case and proposition in FACE. Cells are marked \fcolorbox{white}{gray!36}{$\bigtimes$} for proposition--case pairs with strong evidence and \fcolorbox{white}{gray!18}{$\diagdown$} for pairs with only some evidence.}
    \setlength{\tabcolsep}{1pt}
    \renewcommand{\S}{\cellcolor{gray!40}{\Large $\bigtimes$}}
    \renewcommand{\s}{\cellcolor{gray!20}{\raisebox{1pt}{\large $\diagdown$}}}
    \newcommand{\x}{}
    \centering
    \renewcommand{\arraystretch}{1.2}
    \begin{tabular}{@{}l@{~}l@{\hspace{12pt}}cccccc@{}}
        \toprule
        &&\multicolumn{6}{c}{\textbf{Proposition}}\\
        \cmidrule{3-8}
        \multicolumn{2}{@{}l}{\textbf{Case}} & \textbf{P1} & \textbf{P2} & \textbf{P3} & \textbf{P4} & \textbf{P5} & \textbf{P6} \\
        \midrule
        A & E-commerce algorithms   & \S & \S & \x & \S & \s & \S \\
        B & Local search            & \s & \S & \s & \x & \x & \s \\
        C & E-commerce consultants  & \x & \s & \x & \S & \s & \s \\
        D & Video Streaming         & \S & \S & \S & \S & \S & \S \\
        E & Web shop                 & \S & \s & \s & \x & \x & \s \\
        F & Customer relations      & \x & \s & \x & \S & \s & \S \\
        G & Engineering tools       & \S & \S & \S & \x & \x & \S \\
        H & Web shop \phantom{\Large $\bigtimes$} & \s & \s & \x & \s & \x & \x \\
        I & Web shop                 & \S & \S & \S & \x & \x & \s \\
        J & Product information     & \x & \S & \s & \x & \x & \s \\
        K & Business intelligence   & \s & \S & \x & \S & \x & \S \\
        L & Employee management     & \s & \s & \x & \S & \x & \s \\
        \bottomrule
    \end{tabular}
    \label{tab:empirical_underpinning}
\end{table}

\subsection{CE processes and infrastructure at the case companies (P1: C1$\rightarrow$C2)}
\label{sec:results_p1}

Companies in the study conduct and depend on CE to different degrees (see Table~\ref{tab:companies}). The case companies also have matching degrees of processes and infrastructure support to match their level of CE (with some exceptions discussed in this section). All interviewees at companies with frequent CE mentioned efficient \textit{process and infrastructure} (\textbf{C1}) as crucial for \textit{efficient experimentation} (\textbf{C2}). The analysis on \textbf{P1} is divided into three parts: (1) impact of efficient processes, (2) impact of efficient infrastructure, and (3) the impact of low throughput.

The four companies with High CE expertise in Table~\ref{tab:companies}, cases A, D, G, and I, also have \emph{efficient processes and infrastructure} (\emph{C1})to support their CE. These cases have a data infrastructure, CE platforms built in-house, all relevant roles, and have established processes for CE. \emph{Case A} has low demand on experiment throughput and instead focus their efforts on advanced statistical techniques and speed of CE. \emph{Case D} and \emph{G} do experiments on all their parts of their software products and have both advanced techniques and a streamlined processes for CE. \emph{Case I} has similar advanced CE at some teams in the company. As expressed by Interviewee I4 who's team conducts large amounts of CE: \emph{``If you have a team that has [CE] in its soul, then you want a streamlined process for how to conduct experiments.''} Developers in \emph{Case I} should be able to put an idea under test within hours of its inception and have experiment results a few days later. They struggled initially with their commercial-off-the-shelf CE platform because the data volume overwhelmed it, until they built their own.

Two case companies in the study, \emph{Case E} and \emph{I}  in particular, struggled with implementing the \emph{processes and infrastructure} (\emph{C1}) with the efficiency that they desired. \emph{Case E} has an ad-hoc software engineering process which causes issues for their CE. According to Interviewee E1, there is a constant change of direction from management in prioritization and what metrics to optimize for, incorrect CE execution such as stopping experiments too soon or not using statistical tests, and management not taking account of CE results. This often led to \textit{ineffective experimentation} (\textbf{C2}). Some of these issues could also be attributed to organizational culture, but the issues manifested as a lack of adherance to \textit{processes} (\emph{C1}).
At \emph{Case I} two of the teams (responsible for search engine and recommendation engine, respectively) achieved \textit{efficient CE}  (\textbf{C2})  by having their own purpose built CE platform for their part of the product, as described above. The other software engineering teams at \emph{Case I}, about 50, used a central support team with a data science focus that conducted the CE independently. This was described as \emph{``totally unreasonable''} by I3 due to the low CE throughput, caused by both technical limitations in the commercial cloud-based CE platform (\emph{C1}) and the overhead of having to involve another team.

The amount and quality of available user data was a frequent topic of discussion for all cases. User data is not an explicit part of FACE, but there are processes for adapting CE to low availability of users data. Both the amount and quality of user data has a direct impact on the \emph{CE throughput} (\textbf{C2}). A certain pre-determined number of data points needs to be collected to get some degree of certainty in the results. Low quality on user data lowers the information that can be learned from each data point, thereby also lowering \emph{CE throughput} (\textbf{C2}). User data is a limitation at companies with vast amounts of user data too, since some experiments only target certain users (e.g., that has a certain feature enabled or belongs to a certain user segment).

\subsubsection{Impact of efficient processes}

CE requires following a rigid processes to ensure trustworthy results. This is important for both qualitative pre-deployment experiments and quantitative post\-/deployment experiments. In qualitative experiments there is much manual and subjective work that must be conducted consistently across users and experimenters. For quantitative experiments the tools must be used correctly to get accurate numbers. In addition, there are many methods and techniques that can be used to get more efficient experimentation:
\begin{itemize}
    \item Having \emph{data-driven prioritization} ensures that the most important experiments are done first (done at \emph{Case A, D, G, I, and J}). This could be done by analysing historic user data to see where users have issues, by surveying users, or by  splitting \emph{changes} into the smallest possible implementation (such that it can be aborted early in case of failure).
    \item Applying \emph{data mining} after performing experiments can be used to get more information out of the experiment results, such as dividing the users into segments and analyzing whether the results differ in the segments (done at \emph{Case A and G}).
    \item Making \emph{power calculations} to figure out how many data points are needed in an experiment is a recommended step to do before the experiment is started~\cite{kohavi2009controlled}. However, all interviewees except the ones at \emph{Case G} admitted to never doing it or using the same calculation for all experiments.
    \item Using both \emph{qualitative and quantitative experiments} is recommended. Many companies only have infrastructure for one type of experiment, but \emph{Case D and I} have both and these companies are able to select the experiments that best suit the situation at hand. 
    \item Optimizing the \emph{statistical test} to a more precise version that suits the given situation increases chances of obtaining statistically significant results (done at \emph{Case G and I}). However, Interviewee G3 also warned about spending too much time on this since it requires specialized knowledge and is technically challenging.
    \item Using \emph{experiment designs} with multiple variables achieves more nuanced results, such as multi-variate tests or multi-armed bandits (done at \emph{Case A and G}). This also requires specialized knowledge and is technically challenging. 
\end{itemize}

\subsubsection{Impact of efficient infrastructure}

Getting started with CE was not described as technically hard by any of the interviewees. As phrased by Interviewee G3, \emph{``The original CE system was like 15 lines of Scala, it was just really really simple. It's interesting how you can start up really easy.''} Also, the requirements for pre-deployment experiments is even lower because the qualitative methods do not rely on technical infrastructure. However, as post-deployment CE is scaled up the demands on \emph{infrastructure} (\textbf{C1}) increase to keep up with having \emph{efficient experimentation} (\textbf{C2}).

When the case companies discovered the value of CE they steadily increased their frequency of CE to cover more of their new developments and also on old features that has never been subjected to experiments. The \emph{infrastructure} (\textbf{C1})is a bottleneck to enable increased CE but all case companies except \emph{Case H} were able to keep up with increased infrastructure demands due to CE being prioritized in the organizations. \emph{Case H} lack the resources needed to increase their infrastructure.

The overall \emph{CE infrastructure} (\textbf{C1}) was similar at all cases albeit at different levels of maturity. The infrastructure includes three parts. 
(1) \emph{Data infrastructure} to store and provide query support for user data. This includes telemetry of user data and product data in all parts of the software, collecting various information about users for segmentation, and a system for storing and accessing this data (i.e., a data warehouse). In addition, all data needs to be stored securely and in compliance with legislation (i.e., GDPR).
(2) A \emph{CE platform} with support for starting and stopping experiments, configuring which metrics to target and what additional metrics will be monitored, support for segmentation and arranging metrics in hierarchies if there are too many, alerting in case things go wrong, ways of running experiments in parallel (non-overlapping in case their changes are conflicting), etc.
Finally, (3) \emph{competences} to develop and support infrastructure and to conduct experiments (see Section~\ref{sec:background_models}). Also, the developers need to be educated in CE if they are to take part in it, which was a challenge for the cases where CE was scaled up, i.e. \emph{Case D, G, and I}.

Of the three infrastructure parts, \emph{data infrastructure} is the most demanding to develop, according to interviewees at \emph{Case D, G, and I}, because data infrastructure must be implemented in the entire software offering, rather than as a standalone development. Investment in data infrastructure is one of the reasons that \emph{Case A} is able to conduct CE with \textit{High expertise} despite their small company and software department size. The case company's product relies on user data for algorithms, such as recommender systems that need the same infrastructure, and the case company was able to use this infrastructure for conducting advanced experiments (multi-variate tests and multi-armed bandit variants, see~\cite{ros2018continuous2}) even though CE is not widespread at the company.

\subsubsection{Impact of low throughput on experimentation efficiency}

There are additional consequences of low throughput to the impact and \emph{effectiveness of experiments} (\textbf{C2}), caused by either low capacity or low speed as follows, besides that companies fail to perform the desired amounts of CE.

A \emph{low capacity} to run experiments increases the risk of releasing features with user-related issues, since only a subset of the desired experiments can be conducted. It is not always obvious what changes will have an impact on users or not. This was something all case companies experienced except the most advanced ones (\emph{Case G and I}). As phrased by Interviewee A1: \emph{``A/B tests are always unpredictable, that has been proven. Again and again by us. We do release some non A/B tested functionality and I can't be totally sure it is all good, but it is what it is. [...] We always focus on what customers need as much as possible and do our A/B tests on that.''} Low capacity is primarily caused by lack of user data or developer resources.

\emph{Low speed} of CE can be frustrating, as Interviewee E1 put it when queried about challenges of CE: ``\emph{long lead times make the whole being data-driven thing almost impossible}''. We identified the following two additional consequences of low speed of CE. (1) Low experiment speed can result in other changes made to the product or market causing the invalid CE results. Due to, for example, changes in company priorities (mentioned by E1 and J1), the experimenter forgetting details about the change (mentioned by A3 and D1), or conflicting software changes that make the change incompatible (mentioned by A3, A4, and K2). In addition, (2) it will be hard to use the insights to form hypothesis in the follow-up projects if they are started before the previous projects' experiment is completed. Thus, CE is not really \emph{continuous} when experiments are slow (mentioned by B1).

The issues with lack of speed were observed at some of our case companies though caused by different factors. At \emph{Case A} the lack of speed was due to challenges with continuous delivery and at \emph{Case B} it was due to having experiments with very large scope. At \emph{Case D and J} the lack of speed was caused by various inefficiencies in their process due to their recent start with CE. Finally, at \emph{Case C and H} low user data volumes required long lead times for experiments, and at \emph{Case K and L} the use of primarily qualitative methods with much manual work caused lack of speed.

\subsection{CE impact at the case companies (P2: C2$\rightarrow$C3 and P3: C2$\rightarrow$C4)}
\label{sec:results_p2_p3}

All cases have experiences with using CE to increase \textit{problem--solution fit} (\textbf{P2}). However, not all companies are able to affect \textit{product--market fit} (\textbf{C4}). Only \emph{Case D, G, and I} experiment with product--market fit regularly. \emph{Case B, E, and J} target product--market fit only to some extent, \emph{Case B and J} have multiple user groups of which they are only able to target one with experiments, and \emph{Case E} have issues with \emph{CE processes} (\textbf{C1}). The remaining cases do not target product--market fit due to \emph{high problem complexity} (\textbf{P4}) and lack of \emph{incentive structures} (\textbf{P6}).

\emph{Sales metrics} (revenue, conversion rates, churn rate, etc.) were mentioned often in the interviews as an appealing metric to use for CE. As explained by Interviewee D1, such metrics are both directly relevant to business, and users presumably only pay for software that they think fulfills a need for them. As such, when an experiment can use sales metrics, CE can be used to optimize the product for both business needs and user needs simultaneously, thereby increasing \emph{experiment effectiveness} (\textbf{C2}) and impact on \emph{product--market fit} (\textbf{C4}).

According to \textbf{P3}, \emph{quantitative experiments can increase product\--market fit}, while both pre\-/deployment prototyping experiments and post\-/deployment \emph{experiments can increase problem\--solution fit} (\textbf{P2}). None of the case companies were able to target \emph{product\--market fit} (\textbf{C4}) with anything other than controlled experiments with sales metrics. Presumably due to that it is harder to measure \emph{product\-/market fit} in a way that cannot be done with only a prototype since it requires users to actually pay for something, hence sales metrics.

\subsection{Problem complexity at the case companies (P4: C5$\rightarrow$C2)}
\label{sec:results_p4}

When the the product or service aims to solve a \textit{complex problem} (\textbf{C5}), it is challenging to \textit{experiment efficiently} on it (\textbf{C2}). The problem complexity of the offerings at the case companies span a wide range, from low complexity (\emph{Case D and G}),  medium (\emph{Case A and K}), to high (\emph{Case C and L}). By problem complexity, we mean that there are various challenges in the way the offering delivers value to users. We identified three issues with how problem complexity affects CE: (1) problem complexity makes changes hard, (2) complex user experience makes measurements hard, and (3) configurability addresses problem complexity but splits CE effort.


Some degree of \textit{complexity in the user problem} (\textbf{C5}) can be overcome by using qualitative methods, e.g., user interviews, observations, focus groups, etc. User observations are regularly used at \emph{Case K and L} and can be used even when the user experience cannot be quantified into meaningful sessions nor specified as an adequate metric. The CE process with user observations at \emph{Case K and L} is slightly different than for a quantitative controlled experiment. There is usually no control group, instead the experimenter selects a small group of users and observe their interaction with the change in the product and compares with earlier results (as in a natural or quasi-experiments). As such, qualitative methods can evaluate a change and is considered to be a valid experiment. However, the efficiency of qualitative experiments in terms of reliability per work hour is low due to the cost of interviewing, recording, coding, etc., compared to the cost and scalability of a quantitative experiment once the infrastructure is in place. The ability to be precise in exactly which change has what effect is also lower compared to quantitative experiments. Qualitative methods do have advantages with richer data that can be used to explain why changes fail or succeed. But as a method for conducting an experiment with realistic circumstances they are limited.

\subsubsection{Impact of problem complexity on making changes}

Some software offerings can be hard to implement changes on, due to having to be integrated with other software, causing communication barriers with their developers, or that there are requirements or expectations from customers that the software should not change. In such cases, the CE process becomes hard due the the complexity of implementing the change in the experiment. This affects the ability to conduct controlled experiments (\textbf{C2}).

When experimenting with software with integration needs, the experimenter has to interface with software that is external to the organization. The experimenter cannot make changes incompatible to it, or they have to communicate with the software engineering organization responsible for the integrated system, which would cause delays in CE.
At \emph{Case C} their business is centered around integrating different systems to build web shops, Interviewee C2 mentions this regarding their customers ability to A/B test on their site: \emph{``What customers can change is actually only content [text and images]. [...] Supposedly, at best, the customers can A/B what it looks like [user interface].''} \emph{Case H} is one of their customers that did do their own A/B testing with involvement from Interviewee C1. At \emph{Case A}, where their product is used in other companies' software, the interviewees complain that they cannot \emph{change} how their product is being used (sometimes incorrectly) and that deploying the changes takes considerable time for some customers.

Finally, \emph{user or customer expectations or requirements} can hinder CE because some users might not want changes. At most of the case companies, the users are able to overcome changes easily, so this aspect of \emph{problem complexity} (\emph{C5}) was only discussed by interviewees at \emph{Case D, K, and L}. At \emph{Case D} they mention how their B2B product has customer requirements that mandate the behavior of certain features. It is not possible for them to make changes to these features even if it would be beneficial to their other customers. Of the other B2B companies in the study, only \emph{Case D and F} accept customer requirements to their product in this way. At \emph{Case K and L}, the interviewees mentioned that their users do not want frequent \emph{changes} due to their software being used in a corporate setting where users do not want changes to disrupt their work flow.

\subsubsection{Impact of product complexity on defining measurements for controlled experiments}
The user experience of software with a complex and open-ended work flow cannot be neatly quantified into chunks or summarized with a single metric. Sales metrics can sidestep the issue of defining a user experience metric, but not all companies can use it for experiments directly and it might not be a suitable target for all experiments. \emph{Case F, K, and L} cannot use sales metrics at all due to lack of incentives given by their business model, thus \textbf{P6} is not in play,  and they also have \emph{complex and open-ended user experiences} (\emph{C5}). All of the other case companies mention using metrics that are related to the user experience as a target metric to improve \emph{problem--solution fit} (\emph{C3}). User experience metrics are also monitored on experiments that target \emph{product--market fit} (\emph{C4}) to ensure that business value does not come at the expense of users at all cases able to target product--market fit.

\emph{Case F, K, and L} have similar underlying reasons for having a \emph{complex and open-ended user experience} (\emph{C5}) as explained in the interviews. The software is used throughout the day, such that it cannot be neatly quantified. The goal that users have when they use the software is hard for the software engineers to extract and measure, and the number of features is so large that there might not be enough user data on some features to measure it.

While it is always possible to find \emph{something} to measure in the user experience of all software, such as number of clicks or other user interactions, it is far from certain that optimizing those metrics will transfer to concrete gains. According to Interviewee L2, good user experience metrics should measure whether the user is able to accomplish their goals with using the software or not. Clicks was specifically mentioned as often being a ``\emph{terrible metric}'' by Interviewee A4 and G3 due to being able to be too easily influenced by experiments without having a real impact on more relevant measures. Intuitively, the number of clicks should be kept low to have an efficient user experience. But if the program is something users enjoy spending time on (and revenue is earned through that), then the number of clicks will be beneficial to increase instead. 

\subsubsection{Impact of configurability on CE}

When software is built to be configured into multiple unique variants (as in software product line development~\cite{pohl2005software}) the problem of having enough user data for each feature is exasperated. In such cases, the amount of users for each such unique configuration is lower than for the total set of users, which leads to experiments taking longer to complete due to having less amounts of user data per time period. Having software be capable of configuration is viewed here as a consequence of dealing with high problem complexity. \emph{Case A} has a product with some degree of configuration. They do experiments at different customers and analyze them independently. However, they have a limited number of customers so it is not described by the interviewees as very challenging for them to handle. \emph{Case F} has a product with a very high degree of configuration and they are not able to conduct CE partially because of that. They have a lot of features in their product and the features exist in multiple variants and experiments would have to be repeated for the different variants, which is not efficient or might not be feasible.

\subsection{Business model pivots at the case companies (P5: C6$\rightarrow$C5)}
\label{sec:results_p5}

According to \textbf{P4}, there are limits to what can be done with CE when \emph{the problem complexity} (\textbf{C5}) is high. Companies can simplify what problem the software solves by pivoting their business model to solve another problem, one for which CE can be applied to achieve a higher \emph{problem--solution fit} (\textbf{C3}). CE is not suitable for all development tasks~\cite{bosch2018takes,melegati2019hypotheses} and is unlikely to help reduce problem complexity significantly; we see two reasons for that. First, while CE can improve \emph{problem--solution fit} (\textbf{C3}), actually changing what \emph{problem} the software solves is a large \emph{change} that goes beyond the scope of CE. Changes in an experiment need to be sufficiently small such that the change can be reversed in case of bad results. Second, high \emph{problem complexity} (\textbf{C5}) limits the ability to gain from experimentation (\textbf{P4}), which is a catch-22 scenario; CE requires a goal to target, but no goal can be specified due to the high complexity.

\emph{Case D} pivoted recently by changing their target users from business customers to private individuals. After the pivot, their prioritization was data-driven to meet market needs and they had less customer requirements on their user experience, thus they were able to simplify the user experience. They also increased their CE significantly as a result of the pivot. Note that, the pivot that \emph{Case D} conducted was \emph{not} done specifically to enable CE. Rather, the goal was to reach a larger market which needed a simplified product, and increased \emph{experimentation effectiveness} (\textbf{C2}) was a side-effect of simplifying the problem complexity. No other company in the study has gone through a similar pivot, but Interviewee A5, C2, and F1 mentioned pivoting (though not necessarily with that term) in regards to what changes they would have to implement to enable CE in their company.

\subsection{CE incentive structures at case companies (P6: C6$\rightarrow$C4)}
\label{sec:results_p6}

Some companies in the study have business models with \emph{incentive structures} (\textbf{C6}) that enhance \emph{experimentation effectiveness} (\textbf{C2}) and enable targeting improved \emph{product--market fit} (\textbf{C4}). These business models provide incentive structures that enable CE, e.g., by having metrics available that software engineers can use to directly affect the product sales and user growth. Part of the differences in incentive structures can be explained by the \emph{sales-led and product-led growth dichotomy}, see Section~\ref{sec:background_growth}. Companies with a business model with product-led growth rely on the product to obtain customers, while the companies with a sales-led growth rely on a sales department to obtain customers. 
Incentive structures impact CE in three ways as explained below: (1) user\--business alignment, (2) sales \& marketing interplay, and (3) increased user-data.

Not all companies' offering revolve around a product or service, e.g. e-commerce companies, and those companies cannot be placed in either category of growth. From a software engineering perspective, the companies providing web shops (\emph{Case E, H, and I}) are somewhat in between product-led and sales-led. While these companies can use sales figures in experiments, \emph{changes} in the software have a limited ability to affect the sales of these web shops. \emph{Case B and J} have multiple user groups with different sales and growth strategies for either role, placing them also in between sales-led and product-led. As such, \emph{Case A, C, F, K, and L} have a sales-led growth, \emph{Case B, E, H, I, and J} have neither, and \emph{Case D and G} have product-led growth.

\subsubsection{Impact of user-business alignment}

According to \textbf{P3}, affecting \emph{product--market fit} (\textbf{C4}) relies on using \emph{quantitative experiments} (\textbf{C2}) that target sales metrics. While the availability of sales metrics is necessary to experiment on \emph{product--market fit} (\textbf{C4}), \emph{incentive structures} (\textbf{C6}) are also necessary. \emph{Case A and C} are examples that have sales metrics from their B2B customers, but cannot target \emph{product--market fit} (\textbf{C4}) with those figures since an increase in their customers sales figures does not come with direct business value to them. As such, CE to improve the sales figures is only done for \emph{product--market fit purposes} (\textbf{C4}) at \emph{Case A} and not at all at \emph{Case C}.

In contrast, \emph{Case D and G} use various sales metrics extensively in experiments, presumably due to their user-business alignment. At both cases, the sales metrics come from software sales and they also both have a subscription-based license model where the user pays continuously to use their service. Interviewee G3 phrased it as such: \emph{``Our incentives are mostly aligned with our customers' because it wouldn't mean anything if somebody purchases [our flagship product] and then decide that they hate it and a week later they cancel the subscription.''} Interviewee D2 also expressed the importance of user-business alignment: \emph{``So the way you build things for freemium is that it has to solve a user problem, that's part of the product breed, that's part of why you do the thing, all your metrics align to that.''} Freemium is a license variant where the product has a free and a paid premium version, see Section~\ref{sec:background_growth}. 

\textit{Case D and G} both also use metrics from the sales funnel to find and prioritize issues with their product. As explained by Interviewee D2: \emph{``There's industry standard metrics, like the AARRR model Acquisition, Activation, Retention, Revenue, and Referral. We use them to analyze our platform and our product and see where we are missing things.''} The AARRR Pirate Metrics Framework was proposed by McClure~\cite{mcclure2007pirate} and covers multiple business model aspects.

\subsubsection{Impact of development and sales \& marketing interplay}

Companies with product-led growth models will have the same \emph{incentives} (\textbf{C6}) on the software engineering department and the sales \& marketing department: to increase the sales figures. This makes it easier for them to cooperate. At \emph{Case D} they had a few individuals with the hybrid role of growth engineer that worked as an intermediary between the two departments. At \emph{Case G} they included growth marketers in their specialized team that conducted experiments. The growth engineers/marketers were primarily tasked with finding and analyzing hypotheses about the product to experiment on.

\subsubsection{Impact of increased user-data}
\label{sec:results_p6_user_data}

The case companies that have a strategy of finding many customers and growing rapidly have \emph{incentive structures} (\textbf{C6}) that lead to gaining access to more data as the company's user base increases. User-data volume is critical for CE throughput and to affect \emph{problem--solution fit} (\textbf{C3}) and \emph{product--market fit} (\textbf{C4}). The interviewees at cases with sales-led growth (\emph{Case A, C, F, K, and L}) all mentioned a focus on large important business customers that can pay more instead of many customers, which gives a higher return on the manual work that the sales force put in. Another difference lie in that the product-led growth cases (\emph{Case D and G}) both have a freemium license model that further increases user-data, since the users that use the software for free contribute with user-data.

\section{Discussion}
\label{sec:discussion}

We have presented a theory of Factors Affecting Continuous Experimentation (FACE) on what contextual factors in a company's business model and software organization affects continuous experimentation (CE) and how. The theory is based on empirical observations in 12 case companies through semi-structured interviews and subsequent theory building. In summary, the theory states that processes and infrastructure increase experiment throughput (\textbf{P1}) and experiments can increase problem-solution fit (\textbf{P2}) and quantitative experiments can increase product-market fit (\textbf{P3}). However, CE is limited by how complex of a problem the software solves for users (\textbf{P4}), the problem complexity can be reduced by pivots in the business model (\textbf{P5}). Finally, CE on product-market fit requires business models with the right incentive structures that connect business and user value (\textbf{P6}).

There are two discussion points: (1) what the limitations of CE as a method for software engineering are and (2) which factors affect a company's ability to conduct CE. 

\subsection{What are the limitations of CE?}
\label{sec:discussion_limits}

There are \emph{limitations to what can be achieved with CE}. According to best practice in CE (see Section~\ref{sec:background_ce}), each experiment should be split into the smallest constituent to avoid having to do unnecessary implementation work in case the change is bad. As such, CE entails incremental work. In the road-map for continuous software engineering, Fitzgerald and Stol~\cite{fitzgerald2017continuous} argue that sometimes abrupt changes are needed when creativity and innovation is involved, because incremental work constraints the creativity to similar solutions to what currently exists. 

Furthermore, in CE, experiments are ultimately used as a method for optimizing software towards a given goal by taking small steps in the right direction. However, there is a risk of getting stuck in local optima where no individual small change can improve the current software design, but there might be a better solution to be found if a larger change is introduced. Larger changes are pivots, the impact of which in FACE are described by the proposition \textbf{P5}. When a pivot is introduced to a product that has been polished with CE, it is quite likely that the new version of the product performs worse because it is less polished. There could be minor problems that over time could be improved with CE to find a better solution. Whether or not it is worth the effort of maintaining multiple versions of the software product during this period is ultimately a business decision that cannot be settled within the confinements of CE.

The value provided by CE diminishes over time. When CE is first introduced at a company there are low-hanging fruits to experiment on. At the case companies, A, D, E, G, I, and J, there were long held assumptions about the product and customers that were finally able to be tested and this led to some surprises. However, over time CE becomes established at the company and more parts of the product are stabilized. Also, the size of experiments becomes smaller since the experimenters become more adept at reducing the scope of the experiments, which has also been reported previously~\cite{kohavi2013online}. Hence the utility of each individual experiment tends to decrease unless a pivot happens.

Finally, FACE cannot make claims about applicability of CE to companies outside the defined scope. However, we argue that an experiment-driven approach to software development is unsuitable for much of the software outside the scope, that is, software that is not user-intensive (e.g., company intranet websites) or not user-facing (e.g., low level software libraries). Other requirements elicitation techniques or software performance optimization methods might be better suited for these examples, such as using profiling tools to pinpoint performance bottlenecks with low lead time.

In summary, CE can increase problem--solution fit and product--market fit. However, CE should not be the only strategy for improving software at companies due to the limitations of incremental work. Pivots also play an important role in enabling the product to be optimized with CE.

\subsection{What factors are at play in CE?}
\label{sec:discussion_gains}

In this study, we show that CE is used for different purposes: problem--solution fit or product--market fit. Similar observations has been done before. Schermann et al.~\cite{schermann2018we} and Ros and Bjarnason~\cite{ros2018continuous} describe that experiments are either regression-based to verify or validate features, or business-based to optimize. Our results highlight that most companies are only able to experiment with improving problem--solution fit and few are able to affect product--market fit. However, according to the entrepreneur Andreessen~\cite{andreessen2007part}: \emph{``The only thing that matters is getting to product--market fit''}. Our main research goal in constructing the theory is to find the factors that explain differences in why companies can apply CE to different effect. For example, why some companies are able to experiment with product--market fit and others cannot.

From previous research, we know that the availability of user data is the main limitation~\cite{kohavi2009controlled} and that offering software-as-a-service (SaaS)~\cite{kohavi2009online} facilitates easier CE. Also, there is substantial work conducted on the challenges of applying CE in a B2B setting~\cite{rissanen2015continuous,ros2018continuous,yaman2016transitioning} and in cyber-physical systems~\cite{bosch2012eternal,giaimo2016continuous,mattos2018challenges}. FACE synthesizes these findings; in fact most of these studies are all on aspects of complexity in the problem the product solves for users; corresponding to \textbf{P4}. FACE highlights additional factors as follows.

Three of the propositions (\textbf{P1}, \textbf{P4}, and \textbf{P6}) affect companies' ability to use CE to increase problem--solution fit or product--market fit and those form the factors. The first (\textbf{P1}), on CE processes and infrastructure, covers many aspects and not all are regarded by us as very impactful on whether a company can apply CE or not. While the throughput would be lower without sufficient processes and infrastructure, it is unlikely to be a complete blocker to CE. In addition, several of the companies (see Section~\ref{sec:results_p1}) have described how they gradually improved their CE support in parallel with ramping up CE. Data infrastructure is an exception that requires significant investment. The final three \emph{derived factors} correspond to constructs \textbf{C1}, \textbf{C5}, and \textbf{C6}, and are: 
\begin{enumerate}
    \item \emph{Data infrastructure} is needed to be able to use the metrics that are desired and to cover telemetry of all parts of the software. Companies that rely on user data for other parts of their products, such as recommendation systems, will get a head start on CE by reusing data infrastructure. 
    \item \emph{User problem complexity} is the complexity of the problem that the software solves for users. High problem complexity makes quantifying the user experience hard and making changes in the software hard, which severely impacts the ability to experiment. Modifying the user problem is challenging but possible with pivots in the business model.
    \item \emph{Incentive structures} are required to provide a measurable link between business value, user value, and software engineering activities, such that experiments can affect product\--market fit. Companies that do not have access to sales metrics in experiments will derive less benefits from experiments since user experience metrics are hard to define in a way that they can be used for optimization.
\end{enumerate}



\section{Conclusions}
\label{sec:conclusions}

We conducted a multi-case study with 12 companies and built a theory called Factors Affecting Continuous Experimentation (FACE) based on the empirical material for understanding what factors are at play for companies conducting continuous CE. Six propositions are included in FACE. (1) \textit{Efficient infrastructure and process improves experimentation effectiveness}. Starting with experiments is easy, but companies can put endless effort into: scaling up experiments, obtaining more insights from experiments, and improving the speed of experiments. (2) \textit{Experiments can affect either the problem--solution fit} or (3) \textit{product--market fit}. Targeting product--market fit is far more challenging, but preferable, since it can improve business and user value simultaneously. Many companies are restricted to only improve the user experience and thereby problem--solution fit. (4) \textit{The complexity of the problem the software solves for users strongly limits experiment applicability}. High complexity can limit the ability to make desired changes in the software or to quantify the user sessions. Following that, (5) \textit{pivots in the business model is necessary to simplify the problem complexity}, experiments on their own are unlikely to succeed. Finally, (6) \textit{improving product-market fit needs incentive structures} in the form of metrics from the sales process.

FACE can be used to evaluate company contexts to gauge CE applicability at their companies. There is still future work to further validate the theory by using it in this way to evaluate companies. Additionally, we plan to derive guidelines from the theory to further guide practitioners in their CE.


\subsection*{\textbf{Acknowledgements}}

We want to thank all the participating anonymous companies and  interviewees for their contribution to this project. Thanks to Klaas-Jan Stol, Helena Holmström Olsson, Lars Bengtsson, and Fabian Fagerholm for their feedback on the manuscript version in the first author's PhD thesis. This work was partially supported by the Wallenberg Artificial Intelligence, Autonomous Systems and Software Program (WASP) funded by Knut and Alice Wallenberg Foundation.

\bibliographystyle{spmpsci}
\bibliography{main}

\appendix
\section{Interview Guide}
\label{sec:guide}

The following questions should be adapted to suit the interviewees background and role. The nested bullet lists indicate probes that are only asked when the answer to the main question requires clarification.

\subsection{Introduction}
\begin{itemize}
    \item Inform about consent
    \begin{itemize}
        \item Interview will be recorded
        \item Free to withdraw at any time (including afterwards)
        \item Data will be treated confidentially and anonymized
    \end{itemize}
    \item Explain purpose of study
    \begin{itemize}
        \item Investigating context and process of experimentation and data use
    \end{itemize}
\end{itemize}

\subsection{Case context}
\begin{enumerate}
    \item What does your company do?
    \item What is the overall business strategy?
    \item What is the business model? 
    \begin{enumerate}
        \item What user problem does your offering solve?
        \item Solution: Product? Service? Consultancy? Other?
        \item Who are the customers? B2B? B2C? B2X?
        \item What are the key metrics and how is it measured (channels)?
        \item Costs and revenue?
    \end{enumerate}
    \item How stable is the business model?
    \begin{enumerate}
        \item How do you know when to pivot (change direction)?
    \end{enumerate}
    \item How many employees are there in: Total? Dev? Sales \& marketing? Ops?
    \item Could you describe your own role(s)?
    \begin{enumerate}
        \item What is your background?
    \end{enumerate}
    \item What does your team do within the company?
    \begin{enumerate}
        \item Is it cross-functional?
    \end{enumerate}
    \item In what ways does your company use user data?
    \begin{enumerate}
        \item Is there a specialized data science or engineering team?
        \item Could you shortly describe your overall infrastructure for data?
    \end{enumerate}
    \item How does your SE team prioritize what to build?
    \item How does user data help with prioritization?
\end{enumerate}

\subsection{Experimentation process}
\begin{enumerate}
    \item How was experimentation introduced at the company?
    \begin{enumerate}
        \item In what department was it started?
        \item In what departments is it done now?
    \end{enumerate}
    \item Why do you do experiments?
    \begin{enumerate}
        \item Knowledge? Prioritization? Optimization? Regression? Validation?
    \end{enumerate}
    \item Could you break down the steps that are taken by you and your team when conducting an experiment?
    \begin{enumerate}
        \item Duration? Roles? MVF? Analysis? Power?
        \item Are you aware of any missing steps?
    \end{enumerate}
    \item What do you experiment on?
    \item To what extent do you experiment?
    \begin{enumerate}
        \item Duration? Frequency? Coverage?
    \end{enumerate}
    \item Is there overlapping experimentation?
    \begin{enumerate}
        \item From different teams?
        \item Coordinated?
    \end{enumerate}
\end{enumerate}

\subsection{Experimentation details}
\begin{enumerate}
    \item Could you describe your infrastructure for experimentation?
    \begin{enumerate}
        \item CI/CD pipeline? Reporting? Reliability? Scalability?
        \item Do you use blue/green deployment or feature flags?
        \item What would you like to improve in your infrastructure?
    \end{enumerate}
    \item What type of experiment designs do you use?
    \begin{enumerate}
        \item A/A experiments? MVTs? Bandit testing?
        \item Do you have decision algorithms taking causal decisions?
    \end{enumerate}
    \item Do you use any qualitative methods?
    \begin{enumerate}
        \item Focus groups? User studies?
        \item Do you have specialized teams or individuals for it?
        \item How does it interplay with quantitative experimentation?
    \end{enumerate}
    \item What types of metrics do you use?
    \begin{enumerate}
        \item What metrics would you want to use?
        \item How does your metric translate to company or team success?
    \end{enumerate}
    \item Are experiments analyzed or executed in different segments?
    \begin{enumerate}
        \item Explicit segments: Verticals? Products? B2B customers? Web pages?
        \item Implicit through data mining?
        \item How do you handle diverging results?
    \end{enumerate}
    \item Do you ever do experiments involving external code bases?
    \begin{enumerate}
        \item How did it affect experimentation?
    \end{enumerate}
    \item How do you share knowledge from experiments?
    \begin{enumerate}
        \item Mail? Meetings? Documentation?
    \end{enumerate}
    \item Do you do any long term follow ups on experiments?
    \begin{enumerate}
        \item Repetitions? Long-running experiments?
    \end{enumerate}
\end{enumerate}

\subsection{Holistic experimentation view}
\begin{enumerate}
    \item What are the main challenges with experimentation?
    \item What are the main benefits with experimentation?
    \item Do you face any ethical dilemmas involving your use of user data or experimentation?
    \item How do you strive to improve your experimentation?
\end{enumerate}

\subsection{Final remarks}
\begin{enumerate}
    \item Do you have any final comments, anything that should have been asked?
    \item Could you recommend us any additional interviewee (or organization?)?
    \item We will get back to you within 1--2 weeks about a summary of what was said here.
\end{enumerate}

\section{Code Book}
\label{sec:codes}

All paragraphs of the text were coded, multiple codes can be used on the same paragraph. Some codes are marked with an X to indicate that they can vary, for example Scenario X should be used as Scenario optimization or Scenario validation, these were expanded freely during coding. The division of detail codes into sections are intended only for improved readability, they are not themes or categories. The first two sections indicate context around experimentation, and the last two are on experimentation process and evidence of actual use.

\subsection{Software development and infrastructure}
\begin{itemize}
    \item Testing (\emph{any testing aspects discussed})
    \item Prioritization (\emph{any prioritization aspects discussed}) 
    \item System architecture X (\emph{description of type of architecture e.g. embedded or micro services})
    \item Data infrastructure (\emph{description of e.g. data warehouses, query engines, data science teams, data engineering teams})
    \item Data availability (\emph{whether data is capable to be used for experiments})
    \item Data governance (\emph{cleaning, data provenance, data})
    \item Experimentation platform (\emph{description of an experimentation platform in use at company})
    \item Infrastructure improvements
    \item Infrastructure challenges
    \item Organizational structure (\emph{How departments are structured and how much of e.g. sales and development there is at a company with relevance to experimentation})
    \item Company culture (\emph{descriptions of culture that influences experimentation})
    \item Knowledge sharing (\emph{how knowledge is shared between departments })
    \item Technical infrastructure
    \item Software stack technology
    \item CICD pipeline
    \item Infrastructure maturity
\end{itemize}

\subsection{Business model and strategy}
\begin{itemize}
    \item Pivoting (\emph{change in business model according to strategy})
    \item Product customization (\emph{general or tailored to different market segments})
    \item Product complexity (\emph{description of a complex product in e.g. size or user experience})
    \item Problem-solution pair (\emph{what problem the product solves for users})
    \item Key metrics
    \item Market constraints (\emph{ethics, legislation})
    \item Cost structure
    \item Revenue stream (\emph{pricing model})
    \item Growth model (\emph{how new customers are acquired, channels, customer segments, number users, etc.})
    \item Channels (\emph{path to customers})
    \item Unique value proposition
    \item Target customers
    \item Unfair advantage
\end{itemize}

\subsection{Experimentation process}
\emph{These codes describe strategies for various stages in the process and why experimentation is conducted.}
\begin{itemize}
    \item Ideation (\emph{how ideas/hypotheses are elicited and prototyping is performed at the company})
    \item Experiment design (\emph{how design are decided, pros and cons of designs})
    \item Metrics (\emph{which are used, what specific metrics mean, how they are derived})
    \item Analysis (\emph{how analysis is conducted})
    \item Scenario X (\emph{different experiment archetypes e.g. optimization, validation, verification, learning})
    \item Dark patterns (\emph{a dark pattern is an unethical anti-pattern in UX for tricking users})
    \item Experimenter X (\emph{used to indicate what role initiates and owns an experiment})
    \item Role X (\emph{used to indicate experimentation involvement})
    \item Experiment handover (\emph{description of how experimentation is conducted by some specialist and then hand over to product team after completion})
    \item Experiment inhibitor (\emph{something hinders experimentation, use in combination with another context giving code})
    \item Experiment enabler (\emph{something enables experimentation, use in combination with another context giving code})
    \item Experimentation frequency
\end{itemize}

\subsection{Experimentation usage}
\emph{These codes are used to gauge how much experimentation is done at a particular company, the first three should be mutually exclusive at a company unless there is some conflicting reports by different interviewees.}
\begin{itemize}
    \item Experimentation awareness (\emph{awareness of experimentation at a company but no intent to start})
    \item Experimentation intent (\emph{aware of experimentation at a company and want to start experimentation but have not done so yet})
    \item Experimentation adoption (\emph{in the process of adopting experimentation and/or increasing scale of exp})
    \item Experiment duration
    \item Experiment goal
\end{itemize}
\emph{These codes are used when a technique is mentioned as being used at the case.}
\begin{itemize}
    \item Sprint experiment (\emph{experiment is conducted as part of ordinary development and is prioritized with other software development})
    \item Stand-alone experiment (\emph{experiment that is conducted outside of the development organization})
    \item Controlled experiment usage (\emph{A/B test, quasi experiment})
    \item Optimization usage (\emph{multi-armed bandits, A/Bn tests, MVT, simulations})
    \item Qualitative methods usage (\emph{focus groups, observations})
    \item Survey usage (\emph{questionnaires})
    \item Data mining usage
    \item Feature flag experiments (\emph{considerations for running experiments through feature flags in the same deployment environment})
    \item Blue-green deployment experiments  (\emph{considerations for running experiments through parallel deployments})
    \item Repeating experiments (\emph{same hypothesis repeated for some reason e.g. disbelief, bugs, etc.})
    \item Experimentation cycles (\emph{describes an actual cycle of experimentation where one experiment lead to a new hypotheses and so on})
    \item Overlapping experiments (\emph{considerations for running many experiments in parallel})
\end{itemize}

\end{document}